\documentclass[aps,prx,superscriptaddress, twocolumn]{revtex4-2}

\usepackage{graphicx}
\usepackage{amsmath,amssymb,amsthm,mathtools,mathrsfs}

\usepackage{physics}
\usepackage{comment}
\usepackage{url}
\usepackage{color}
\usepackage{xcolor}
\usepackage{graphicx}

\newcommand{\trans}{{\scriptscriptstyle\mathsf{T}}}

\usepackage{bm}
\usepackage{times}

\usepackage[colorlinks,citecolor=teal,linkcolor=teal, urlcolor=teal]{hyperref}

\newtheorem{thm}{Theorem}

\newtheorem{dfn}{Definition}
\newtheorem{lem}[thm]{Lemma}

\newtheorem{cor}[thm]{Corollary}

\usepackage{appendix}

\newcommand*\pFqskip{8mu}
\catcode`,\active
\newcommand*\pFq{\begingroup
        \catcode`\,\active
        \def ,{\mskip\pFqskip\relax}%
        \dopFq
}
\catcode`\,12
\def\dopFq#1#2#3#4#5{%
        {}_{#1}F_{#2}\biggl[\genfrac..{0pt}{}{#3}{#4};#5\biggr]%
        \endgroup
}

\DeclareMathOperator{\arsinh}{arsinh}

\begin{document}

\title{More global randomness from less-random local gates}
\author{Ryotaro~Suzuki}
\email{ryotaro.suzuki.2139@gmail.com}
\affiliation{\fu}


\author{Hosho~Katsura}
\affiliation{\utphys}
\affiliation{\utpi}
\affiliation{\utts}

\author{Yosuke~Mitsuhashi}
\affiliation{\utbasic}

\author{Tomohiro~Soejima} 
\affiliation{\hu}

\author{Jens~Eisert}
\affiliation{\fu}

\author{Nobuyuki~Yoshioka}
\affiliation{\uticepp}

\newcommand{\fu}{Dahlem Center for Complex Quantum Systems, Freie~Universität~Berlin, Berlin~14195, Germany}

\newcommand{\utphys}{Department of Physics, The University of Tokyo, Tokyo 113-0033, Japan}

\newcommand{\utpi}{Institute for Physics of Intelligence, The University of Tokyo, Tokyo 113-0033, Japan}

\newcommand{\utts}{Trans-Scale Quantum Science Institute, The University of Tokyo, Tokyo 113-0033, Japan}

\newcommand{\utbasic}{Department of Basic Science, The University of Tokyo, Tokyo 153-8902, Japan}

\newcommand{\hu}{Department of Physics, Harvard University, Cambridge, MA 02138, USA}

\newcommand{\uticepp}{International Center for Elementary Particle Physics, The~University~of~Tokyo, Tokyo~113-0033, Japan}

\begin{abstract}
Random circuits giving rise to unitary designs are key tools in quantum information science and many-body physics. In this work, we investigate a class of random quantum circuits with a specific gate structure. Within this framework, we prove that one-dimensional structured random circuits with non-Haar random local gates can exhibit substantially more global randomness compared to Haar random circuits with the same underlying circuit architecture. 
In particular, under a solvability condition, we exactly diagonalize the second-moment operators for these structured random circuits by establishing a link to free-fermionic models, and we show that their spectral gaps can exceed those of Haar-random circuits. While solvable structured random circuits exhibit free-fermionic behavior, the underlying fixed gate need not be a matchgate. Our findings have applications in improving circuit depth bounds for randomized benchmarking, in generating approximate unitary 2-designs from shallow random circuits, and in enabling exact analyses of quantum many-body dynamics.
\end{abstract}

\maketitle

\let\oldaddcontentsline\addcontentsline
\renewcommand{\addcontentsline}[3]{}

\section{Introduction}
Randomness is a ubiquitous concept across various subfields of physics, mathematics, and computer science.
In particular, random unitary ensembles have a wealth of applications in quantum information processing, such as the error estimation via randomized benchmarking~\cite{emerson2005scalable, knill2008randomized,PhysRevLett.106.180504, BenchmarkingReview}, providing computational advantage over classical computers~\cite{boixo2018characterizing, arute2019quantuma, Gao_2024, ware2023sharp, morvan2023phase, fefferman2024anti,SupremacyReview}, and verification of the state/process through quantum tomography~\cite{huang2020predicting, zhao2021fermionic, elben2023randomized, bertoni2024shallow, wan2023matchgate}. 
Presumably the most 
commonly considered random unitaries, particularly in theoretical studies, are the global Haar random unitary ensemble, for which 
a unitary is chosen uniformly at random from the entire 
unitary group acting on the target system. 
However, while compelling mathematically, such a construction 
is notoriously impractical for systems of intermediate size.
Consequently, genuine Haar randomness is often sacrificed for practicality by employing a sequence of \emph{local} random circuits, which have their own 
wealth of applications in quantum science, such as modeling quantum chaotic many-body dynamics~\cite{PhysRevX.8.021014, PhysRevX.7.031016, hosur2016chaos, roberts2017chaos, PRXQuantum.2.030316, haferkamp2022linear, suzuki2023quantum, fisher2023random} and exploring the expressiveness of unitary operations~\cite{harrow2009random, BHH_unitary_designs, mittal2023local, belkin2023approximate,  chen2024incompressibility, deneris2024exactspectralgapsrandom, fefferman2024anti, fisher2023random,marvian2022restriction,  mitsuhashi2024unitary, mitsuhashi2024characterization, hulse2024unitary}, as well as the applications for quantum information processing we have raised above.

Building on the seminal work by Brand\~ao, Harrow, and Horodecki, which have shown that local random circuits can approximate $n$-qubit Haar random ensembles up to $k$-th order in $O(nk^{10.5})$ depth under brick-wall architecture~\cite{BHH_unitary_designs}, there has been intensive pursuit to quantify how precisely the local random gate structure affects global randomness ~\cite{PhysRevX.7.021006, RandomHamiltonians,Haferkamp2022randomquantum, haferkamp2022linear, deshpande2022tight, mittal2023local, belkin2023approximate, harrow2023approximate,schuster2024random, laracuente2024approximate,chen2024incompressibility}. Most studies have assumed a fixed geometry of local random unitary gates, showing that
a linear number of Haar random local gates in $n$ can generate approximate unitary $k$-designs~\cite{Haferkamp2022randomquantum,  mittal2023local, belkin2023approximate, harrow2023approximate,gross2007evenly}; recent advances have achieved logarithmic depth scaling with $n$ through gluing of logarithmic-size random circuits~\cite{schuster2024random, laracuente2024approximate,watts2024quantumadvantagemeasurementinducedentanglement}.
Moreover, as a seminal result, a linear growth of approximate design orders has recently been achieved \cite{chen2024incompressibility}.

An outstanding and practically relevant question that arises is: {\emph {Do local Haar random gates actually constitute the best randomizer to achieve global randomness?}}
{Similar questions have also been raised in Refs.~\cite{haferkamp2023momentsrandomquantumcircuits, schuster2024random, harrow2023approximate} as open problems.
While this statement may be intuitive, recent works have suggested that Haar random gates need not be optimal.
Indeed, Gao \textit{et al.}~\cite{Gao_2024} and Ware \textit{et al.}~\cite{ware2023sharp} have numerically demonstrated that Haar random gates do not extremize global randomness in the context of the quantum computational advantage. Earlier works by \v{Z}nidari\v{c}~ \cite{vznidarivc2007optimal,PhysRevA.78.032324} have also numerically shown that the convergence rate of random circuits can be improved by adding a gate structure in the context of entanglement dynamics, and have provided an analytical argument for all-to-all interacting circuits in the large system size limit.}

{In this work, we analytically and rigorously answer the above question by investigating the model of random circuits with non-Haar local random gates, consisting of a fixed interacting gate $u$ sandwiched by single-qudit Haar random gates, which has been previously considered in Refs.~\cite{vznidarivc2007optimal, PhysRevA.78.032324, Gao_2024, ware2023sharp}.}
We quantify the global randomness by the spectral properties of the second-moment operators of structured random circuits. We identify a solvability condition on the gate structure such that we can obtain all the eigenvalues and eigenvectors, not only the spectral gap. 
{As a consequence, we prove} that 
random circuits consisting of structured random local gates can be substantially 
{more random} than those of Haar random local gates.

{A key technique of our work is the mapping of structured random circuits to free-fermionic models, which identifies a gate set with desirable properties and yields new exactly solvable models of quantum many-body dynamics.
We find that the second-moment operators of structured local random circuits correspond to the {anisotropic} $XY$ model \cite{LIEB1961407} under the solvability condition and, without the condition, a frustration-free $XYZ$ chain in a magnetic field \cite{PhysRevB.92.115137, PhysRevB.95.195140, vardhan2024entanglement}.
We exactly diagonalize the second-moment operator by the free-fermionic technique \cite{LIEB1961407}.
Interestingly, these solvable models include fixed {\it non}-matchgates $u$ that behave as free fermions only because of the surrounding single-qudit Haar-random unitaries. Here, matchgates refer to the class of two-qubit unitaries that generate free-fermionic unitary dynamics \cite{10.1145/380752.380785, PhysRevA.65.032325, jozsa2008matchgates}. } 
We also discuss possible applications of our results, such as reducing depth bounds for the {filtered} randomized benchmarking by random quantum circuits \cite{heinrich2022randomized} and the convergence rate to unitary 2-designs in shallow quantum circuits \cite{schuster2024random, laracuente2024approximate, watts2024quantumadvantagemeasurementinducedentanglement, bertoni2024shallow}.

\begin{figure*}[t]
\centering
  \includegraphics[width=16.5cm]{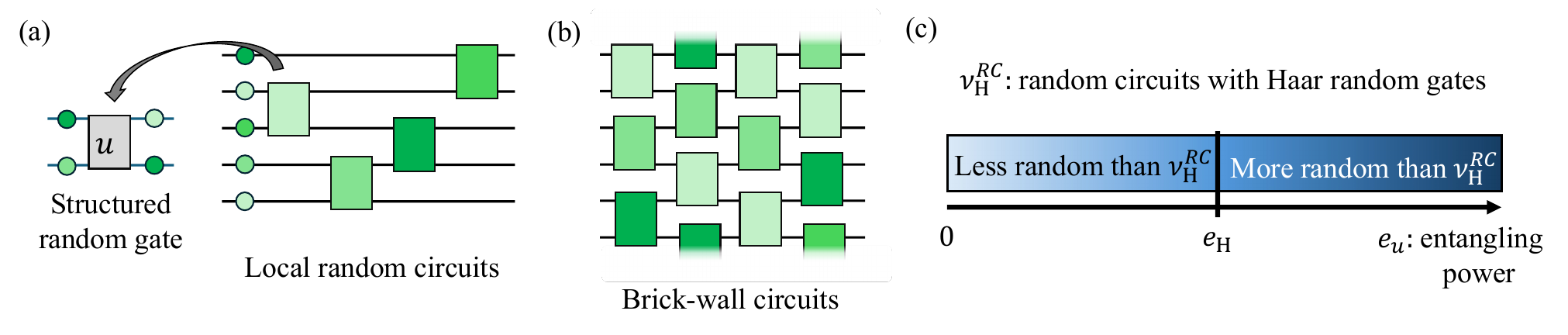}
  \caption{
  Setup and main results. Structured random gates consist of a fixed two-qudit gate $u$ and four Haar random single-qudit gates sandwiching it. We consider two models of random circuits, where structured random gates are applied in the (a) local random circuit architecture and (b) brick-wall circuit architecture, where the boxes are two-qudit gates and the balls are single-qudit Haar random unitaries.  As stated in Theorems~\ref{thm:more_randomness_loca_random} and \ref{more_randomness_brick_wall}, we show that if the entangling power of a structured random gate is greater than that of Haar random gates, the structured random circuits satisfying a solvability condition are more random than Haar random circuits with the same architecture (c).}
  \label{fig:setup}
\end{figure*}

\section{Setup for structured random circuits.} 
We investigate how the precise structure of random local gates affects the global randomness of an entire system, and to this end we introduce models of random circuits with non-Haar random gates. Concretely, we consider a random two-qudit gate generated by fixing a two-qudit gate $u$ and sandwiching it by single-qudit Haar random gates acting on the inputs and the outputs, as shown in Fig.~\ref{fig:setup} (a), which we call {\emph{$u$-structured random gates}}. 
Formally, let $u\in \mathrm{U}(d^2)$ be a fixed two-qudit unitary operation and let $\nu_{\textrm{H}}$ denote the Haar measure on $U(d)$. 
A $u$-structured random gate is a random unitary $U\in \mathrm{U}(d^2)$ generated as follows:
    \begin{equation}
    U = (v_1\otimes v_2)\,u\,(v_3\otimes v_4), \qquad
    v_1,v_2,v_3,v_4 \;{\sim}\; \nu_{\mathrm{H}}.
    \end{equation}

We consider a one-dimensional system of $n$ qudits, where the local dimension is $d$.
A sequence of random gates specifies a probability distribution $\nu$ in the unitary group, which we interchangeably refer to as the random circuit.
We consider the following  circuit architectures of one-dimensional structured random circuits: (1) {\emph{$u$-structured local random circuit}} $\nu_{u}^L$, where we pick a neighboring pair of qudits uniformly at random and apply a $u$-structured random gate, which is illustrated in 
Fig.~\ref{fig:setup} (a),
(2) {\emph{$u$-structured brick-wall random circuit}} $\nu_{u}^B$, where we apply staggered two layers of $u$-structured two-qudit gates on nearest neighbors, which is illustrated in 
Fig.~\ref{fig:setup} (b).
For technical simplicity, we assume that the number of qudits $n$ is even and the system satisfies the periodic boundary condition.
To exclude a trivial case, we also assume that $n>2$.
For a nearest-neighbour pair of sites $(i,i+1)$ and a two-qudit unitary gate $G$, we write $G_{i,i+1}$ for the two-qudit unitary acting nontrivially on qudits $i$ and $i+1$ and as the identity on all other sites.
Formally, we define structured random circuits as follows.
\begin{dfn}[$u$-structured random circuit]
\label{def_circuit_architecture}
Let us consider $n$-qudit one-dimensional systems
$(\mathbb{C}^d)^{\otimes n}$ with even $n$ under periodic boundary conditions, where $n>2$.
Let $\mu_u$ be the distribution of the two-qudit structured random gate with $u$ on $\mathrm{U}(d^2)$.
The local and brick-wall architectures are defined as follows.

\begin{enumerate}
\item \textit{$u$-structured local random circuit.}
A random unitary $U\sim \nu_u^{L}$ is generated by first drawing a site index
$i\in\{1,2,\dots,n\}$ uniformly at random and then sampling a two-qudit gate
$G\sim \mu_u$. One then applies $G$ to the neighbouring pair $(i,i+1)$, that is,
$U := G_{i,i+1}$.

\item \textit{$u$-structured brick-wall random circuit.}
The one-layer distribution $\nu_u^{B}$ on $\mathrm{U}(d^n)$ is defined by
\begin{equation}
U :=
\left(\, \bigotimes_{i:\, \mathrm{odd}} G_{i,i+1}\right)
\left(\, \bigotimes_{i:\, \mathrm{even}} G_{i,i+1}\right),
\end{equation}
where all two-qudit gates $G_{i,i+1}$, for $i\in\{1,2,...,n\}$ are drawn independently from
$\mu_u$ and the index $i$ runs over the odd or even sites of the $n$-qudit chain.
\end{enumerate}
\end{dfn}
\noindent 
For local random circuits, we apply single-qudit Haar random unitaries on all qudits for technical convenience \cite{Note2}. 
We also define local and brick-wall Haar random circuits by $\nu^{L}_{\mathrm{H}}$ and $\nu^{B}_{\mathrm{H}}$, respectively, where we apply two-qudit Haar random gates in the corresponding architecture.

\subsection{Characterization of randomness.} 
Here, we characterize randomness using the moment operator 
\cite{gross2007evenly, harrow2009random}. 
Given a random circuit $\nu$, the second-moment operator is defined by
\begin{equation}
    M_\nu:=\int U^{\otimes 2}\otimes 
 {U^*}^{ \otimes 2} d\nu(U),
\end{equation}
 where $ {U}^*$ is the complex conjugate of $U$.
 We denote moment operators for $u$-structured local and brick-wall random circuits in one step by $M_{L,u}$ and $M_{B,u}$, respectively. For Haar random circuits, they are defined by replacing the subscript $u$ by $\mathrm{H}$. The moment operators of these circuits with $t$ steps, or $t$ depth, are obtained by multiplying them $t$ times, namely $\left(M_{a,u} \right)^t$ for $a=L,B$.
We apply single-qudit Haar random unitaries on all qudits in both architectures, which become the projector onto the $2^n$-dimensional subspace $\mathrm{span}\{ \ket{I}, \ket{S} \}^{\otimes n}$ by the fact that the second-moment twirling map is the projector onto the subspace spanned by the permutation operators \cite{harrow2013churchsymmetricsubspace, BHH_unitary_designs}. 
Here, $\ket{I}$ and $\ket{S}$ are the vectorization of the identity and SWAP operators acting on $\mathbb{C}^{d^2}$.
It is thus enough to consider $M_{a,u}$ restricted to the subspace, for $a=L,B$. 
Later, we show the diagonalization results in this subspace.
We write the eigenvalues of $M_{\nu}$ in non-increasing order in absolute value $|\lambda_1| \geq |\lambda_2| \geq 
 \cdots$, where we have $\lambda_i=1$ for $i=1,2$. 
The largest eigenvalue is $1$, and the corresponding $1$-eigenspace includes the subspace $V_1$ spanned by the $n$-fold vectorization of the identity and SWAP gate, that is,
\begin{equation}
    V_1=\mathrm{span}\{ \ket{I}^{\otimes n}, \ket{S}^{\otimes n}\},
\end{equation}
where $n$ is the number of qudits. 
This subspace $V_1$  turns out to be the entire $1$-eigenspace if a fixed $u$ is an entangling gate, because an entangling gate and the single-qubit gates form a universal gate set \cite{PhysRevLett.89.247902}, and then the $1$-eigenspace of $M_{\nu}$ is equal to that of the second-moment operator of the $n$-qudit Haar random unitary $M_{\mathrm{H}}$ \cite{BHH_unitary_designs}, where the $1$-eigenspace of $M_{\mathrm{H}}$ is exactly $V_1$.

We focus on structured random gates with an entangling gate $u$.
The gap between the largest and second-largest eigenvalues of $M_\nu$ in absolute value, which is $1-|\lambda_3|$, is called the \emph{spectral gap}, denoted by $\Delta_{\nu}$.  
The eigenvalues of $M_{\nu}$ determine the convergence rate to unitary 2-designs
\cite{harrow2009random, BHH_unitary_designs, PhysRevA.80.012304, kaposi2024generalized}, where unitary $k$-designs are ensembles whose $k$-th moment agrees with that of global Haar random unitaries \cite{gross2007evenly}. Concretely, when the eigenvalues of $M_{\nu}$ are small and hence its spectral gap 
is large, we have fast convergence to unitary 2-designs.

We then formalize what ``more randomness'' means in two ways, where the former is based on the spectral gap, and the latter is rooted in the entire property of second-moment operators.

\begin{dfn}[More random with 
respect to spectral gaps]\label{def-more-randomness-gap}
    Let $\nu$ and $\mu$ be probability distributions of random circuits.
    The distribution $\nu$ is said to be {more random} with respect to spectral gap
    than  
    $\mu$ if $\Delta_{\nu}>\Delta_{\mu}$. 
\end{dfn}

\begin{dfn}[More random with respect to positivity]\label{def-more-randomness-positivity}
When $M_{\nu}$ and $M_{\mu}$ are Hermitian operators, 
$\nu$ is said to be {more random with respect to positivity} 
than $\mu$ if the restriction of the operator $\left. M_{\mu} \right|_{V_1^\perp} - \left. M_{\nu} \right|_{V_1^\perp}$ is a positive operator,
where ${V_1^\perp}$ is the orthogonal complement of $V_1$.
\end{dfn}

The spectral gaps of the ensembles forming exact unitary 2-designs are $1$, and, therefore, they are the most random ensembles by Definition~\ref{def-more-randomness-gap}.
We note that $M_{L,u}$ is always Hermitian, as discussed later. 
More generally, when the circuit architecture of $\nu$ is invariant under the time reflection, $M_{\nu}$ is Hermitian.
We note if $\lambda_3 \geq 0$ for both random circuits $\nu$ and $\mu$, and $\nu$ is more random with respect to positivity than $\mu$, we have that $\nu$ is also more random with respect to spectral gap than $\mu$,
where this follows from Weyl's monotonicity theorem \cite{bhatia2013matrix}.

We remark on the definition based on positivity.
One particular feature of Haar random unitaries is that an expectation value is highly concentrated around its mean value \cite{low2009large}. In this context, the bound on the second-moment implies the bound on the variance.
Specifically, more randomness of $\nu$ than $\mu$ with respect to positivity implies that, for a state  vector $\ket{\psi}$, the probability distribution of $| \bra{\psi} U \ket{\psi}|^2$ with $U$ drawn from a structured random circuit $\nu$ is more concentrated around its mean value than that of another structured random circuit $\mu$.
This kind of concentration bound has been 
used, for example, to prove the equilibration of random 
product states \cite{PRXQuantum.2.040308} and 
lower-bound the circuit complexity of the random unitary ensemble \cite{PRXQuantum.2.030316, haferkamp2023momentsrandomquantumcircuits}.

\section{Entanglement and moment operators} 
When the fixed gate $u$ is an entangling gate,
the gate set becomes computationally universal, and eventually, random circuits converge to Haar random unitaries \cite{BHH_unitary_designs, Yada2026}, while when $u$ is not entangling, random circuits do not.
This simple observation suggests that we can parameterize the randomness of structured random circuits by the entanglement property of $u$.
In fact, we discuss that the second-moment operators of structured random circuits depend only on the entangling power $e_u$ 
 \cite{PhysRevA.62.030301, PhysRevA.95.040302,PhysRevResearch.2.043126, PhysRevResearch.3.043034}, which is the linear entanglement entropy of a random product state applied by $u$, and its complementary quantity $g_u$, called the gate typicality \cite{PhysRevA.95.040302, PhysRevResearch.3.043034, PhysRevResearch.2.043126}.
Below, in this section, we first review the entangling power and gate typicality and then 

 \subsection{Entangling power and gate typicality}
 \label{entangling_power_gate_typicality}

One of the common possible ways to define the entangling power of a unitary is as follows: The entangling power $e_u$ of a two-qudit unitary $u$ is quantified by the amount of entanglement of random product states applied by $u$
\begin{align}
    e_u=\frac{d+1}{d-1}\int E_{\textrm{lin}}(u\ket{\psi} \otimes  \ket{\phi}) d\!\ket{\psi}d\!\ket{\phi},
\end{align}
where the integration is over the single-qudit Haar random state and $E_{\textrm{lin}}(\ket{\Psi})\coloneqq 1-\Tr \rho_A^2$, with $\rho_A=\Tr_B{\ket{\Psi}\!\bra{\Psi}}$, is the linear entanglement entropy.  Here, $\ket{\Psi}$ is a state in a two-qudit system, where we have defined the subsystems $A$ and $B$ by the individual qudits. A similar entanglement measure of a unitary $u$ is the operator entanglement entropy. 
Let $\ket{u}$ be the normalized vectorization of a two-qudit unitary $u$,
where $\ket{u}=({1}/{d^2})(u_{A_1,B_1}\otimes I_{A_2,B_2})\sum_{i,j=1}^{d}\ket{i,j,i,j}_{A_1,B_1,A_2,B_2}$, and $\rho_A(u)=\Tr_{B_1,B_2}{\ket{u}\bra{u}}$.
Then, the {operator entanglement} \cite{PhysRevA.76.032316} of $u$ is defined by 
\begin{align}\label{eq-def-operator-entanglement}
    E(u) \coloneqq 1-\Tr \rho_A(u)^2,
\end{align}
and this is equal to
\begin{align} \label{eq-operator-entanglement-rewitten}
    E(u)=1-\frac{\bra{I}\otimes \bra{S} u^{\otimes 2,2} \ket{I} \otimes \ket{S}}{d^4}.
\end{align}
Since $\Tr \rho_A(u)^2$ is the purity of $\rho_A(u)$, the operator entanglement is bounded as $0 \leq E(u) \leq 1-{1}/{d^2}$.
As shown in Ref.~\cite{PhysRevA.62.030301}, the entangling power of a two-qudit unitary $u$ can be rewritten as 
\begin{align}\label{eq-def-entangling-power}
    e_u=\frac{1}{E(\text{SWAP})}\left( E(u)+E(u \cdot \text{SWAP})-E(\text{SWAP}) \right),
\end{align}
where $\text{SWAP}$ denotes the two-qudit swap gate, and it takes the maximum of the operator entanglement as
\begin{align}
    E(\text{SWAP})=1-\frac{1}{d^2}.
\end{align}
Throughout this work, we use $\text{SWAP}$ for the two-qudit SWAP gate acting on one physical system and $S$ for the swap operation acting on two copies of physical  systems.
For two-qubit gate $u$, the entangling power can take all the value within $0 \leq e_u \leq 2/3$, where, for example, the CNOT gate and the iSWAP gate have the maximum value $e_u = 2/3$.
For two-qudit gate $u$ with local dimension $d \geq 3$, the value takes $0 \leq e_u \leq 1$.
For Haar random two-qudit unitary gates, the averaged entangling power is 
\begin{align}
    e_{\mathrm{H}} \coloneq \int e_u d \mu_{\mathrm{H}}(u) =\frac{d^2-1}{d^2+1}. \label{eq-entangling-power-Haar}
\end{align}
For example, in the case of qubit-system, we have $e_{\mathrm{H}}=3/5$.

In addition, the \emph{gate typicality} $g_u$ \cite{PhysRevA.95.040302, PhysRevResearch.3.043034, PhysRevResearch.2.043126}, which is a complementary quantity to $e_u$, is defined by
\begin{align}\label{eq-def-gate-tyoicality}
    g_u=\frac{1}{2E(\text{SWAP})}\left( E(u)-E(u \cdot \text{SWAP})+E(\text{SWAP}) \right).
\end{align}
For two-qudit gate $u$ with local dimension $d \geq 2$, the value takes $0 \leq g_u \leq 1$, where $g_u=1$ and $g_u=0$ if $u$ is the two-qudit SWAP gate and the identity gate, respectively.
For Haar random two-qudit unitary gates, the averaged gate typicality is 
\begin{align} \label{eq_gate_typicality_Haar}
    g_{\mathrm{H}}\coloneq \int g_u d \mu_{\mathrm{H}}(u)=\frac{1}{2}.
\end{align}

We finally remark that there are limitations of feasible $e_u$ and $g_u$, i.e., there does not necessarily exist a unitary gate $u$ for $e_u$ and $g_u$ satisfying the conditions above.
For a two-qubit unitary $u$, the values $e_u$ and $g_u$ are feasible if and only if they satisfy $2g_u(1-g_u) \leq e_u \leq 2g_u$,  $e_u \leq 2-2g_u$, and $e_u \leq \frac{2}{3}$ \cite{PhysRevResearch.2.043126}.
For a two-qudit unitary with $d \geq 3$, the constraint $e_u \leq \frac{2}{3}$ is replaced by $e_u \leq 1$, and the {ranges of $e_u$ and $g_u$ are} numerically shown in Ref.~\cite{PhysRevResearch.2.043126}, see the reference for more details.
In general, for an arbitrary two-qudit gate $u$, we have
\begin{equation}
\begin{aligned}
    e_u \leq 2-\frac{g_u}{g_{\textrm{H}}}, \label{eq:eu-gu-upper-boundary} \\
    e_u \leq \frac{g_u}{g_{\textrm{H}}},
\end{aligned}
\end{equation}
which follow from the definitions of $e_u$ and $g_u$ with the bound $0 \leq E(u) \leq E(\textrm{SWAP})$.
The two-qudit unitary gates satisfying $e_u = 2-2g_u$ are called dual-unitary gates \cite{bertini2019exact, PhysRevB.101.094304, PhysRevLett.126.100603, suzuki2022computational}, and those satisfying $e_u=1$ and $g_u=1/2$ are called perfect tensors \cite{pastawski2015holographic, PhysRevA.92.032316}.

\subsection{Second-moment of structured random gates}
\label{subsec:secomd_moment_structured_gates}
Here, we relate the second-moment operators of $u$-structured random gates with the entangling power and the gate typicality.
The second moment operator of the $u$-structured random gate is  
\begin{align} \label{eq-averaged_fixed_gate}
    W^u \coloneq \int \left( \big(v_1 \otimes v_2 \big) u \big(v_3 \otimes v_4 \big) \right)^{\otimes 2,2} \prod_{i=1}^4d\nu_{\mathrm{H}}(v_i),
\end{align}
where $\nu_{\mathrm{H}}$ is the Haar measure on the single-qudit unitary group $\mathrm{U}(d)$ and we use the notation $U^{\otimes k,k}=U^{\otimes k} \otimes \left( U^* \right)^{\otimes k}$. 
The second-moment operators of $u$-structured random circuits with the local and the brick-wall architectures are written as, respectively, 
\begin{align}
    M^{(2)}_{L,u} &= \frac{1}{n} \sum_{i=1}^n W^u_{i,i+1},\label{def-moment-operator-LRC} \\
    M^{(2)}_{B,u} &=
\left(\, \bigotimes_{i:\, \mathrm{odd}}  W^u_{i,i+1}\right)\left(\, \bigotimes_{i:\, \mathrm{even}}  W^u_{i,i+1}\right),\label{def-moment-operator-BW} 
\end{align}
where $W_{i,i+1}$ is the operator acting nontrivially on sites $i$ and $i+1$ as $W$ and as the identity on all other sites, and by periodic boundary conditions we have that $W^u_{n,n+1}=W^u_{n,1}$. They are direct consequences of the definition of the architectures: the local circuit architecture randomly picks up $i$ and then applies $W_{i,i+1}$, which yields Eq.~\eqref{def-moment-operator-LRC}, and the brick-wall architecture applies $W_{i,i+1}$, for odd or even $i$, which yields Eq.~\eqref{def-moment-operator-BW}.

We can integrate over the single-qudit unitaries in Eq.~\eqref{eq-averaged_fixed_gate} using the Weingarten calculus \cite{collins2006integration}, to get
\begin{equation}
\begin{aligned}
    &\int v^{\otimes 2,2}d\nu_{\mathrm{H}}(v)\\
    &=\frac{1}{d^2-1}\big( 
    \ketbra{I}{I}
    +\ketbra{S}{S}
    -\frac{1}{d}\big(\ketbra{I}{S}+\ketbra{S}{I}\big) 
    \big)\\
    &=\ketbra*{\widetilde{I}}{I}+\ketbra*{\widetilde{S}}{S}
\end{aligned}
\end{equation}
where we have defined
\begin{equation}
\begin{aligned} 
    \ket*{\widetilde{I}} \coloneq \frac{1}{d^2-1}\left(\ket{I}-\frac{1}{d}\ket{S}\right), \\
    \ket*{\widetilde{S}} \coloneq \frac{1}{d^2-1}\left(\ket{S}-\frac{1}{d}\ket{I}\right). \label{eq-def-Itilde-Stilde-states}
\end{aligned}
\end{equation}
The pair of the sets $\{\ket{I},\ket{S}\}$ and $\{\ket*{\widetilde{I}},\ket*{\widetilde{S}}\}$ is a biorthogonal system because they satisfy $\braket*{\sigma}{\widetilde{\tau}}=\delta_{\sigma,\tau}$, where $\delta_{\sigma,\tau}=1$ for $\sigma=\tau$ and $\delta_{\sigma,\tau}=0$ for $\sigma \neq \tau$. 
We also introduce the orthonormal basis
\begin{equation}
\begin{aligned} 
\ket{\overline{0}} \coloneq \frac{1}{\sqrt{2d(d-1)}}\left( \ket{I}-\ket{S} \right),
\label{eq:def_overline_0}\\
\ket{\overline{1}} \coloneq \frac{1}{\sqrt{2d(d+1)}}\left( \ket{I}+\ket{S} \right).
\end{aligned}
\end{equation}
In this basis representation, we have that
\begin{align} 
\int v^{\otimes 2,2}\, d\nu_{\mathrm{H}}(v) \, = 
\,\ketbra{\overline{0}}{\overline{0}} \,+\, \ketbra{\overline{1}}{\overline{1}}. 
\end{align}

With the orthonormal basis $\{\ket{\overline{0}}, \ket{\overline{1}}\}$, Eq.~\eqref{eq-averaged_fixed_gate} becomes
\begin{align} \label{eq-averaged_fixed_gate_Weingarten}
    W^u = \sum_{s_i \in \{\overline{0},\overline{1}\}} W^u(s_1, s_2 ,s_3 ,s_4)
    \ket{{s}_1} \otimes \ket{{s}_2} 
    \bra{s_3} \otimes \bra{s_4},
\end{align}
where we have defined
\begin{equation} \label{eq-def-W-matrixelement}
    W^u(s_1 ,s_2, s_3 ,s_4)=\big(\bra{s_1} \otimes \bra{s_2} \big)u^{\otimes 2,2} \big( \ket{{s_3}} \otimes \ket{{s_4}} \big).
\end{equation}

We now express the matrix elements of $W^u$ in terms of the entangling power and the gate typicality. To this end, by Eqs.~\eqref{eq-def-entangling-power} and \eqref{eq-def-gate-tyoicality}, we calculate that  
\begin{equation}
\begin{aligned}
    E(u) 
    &=\frac{d^2-1}{2d^2}(e_u+2g_u),\\
    E(u \cdot \text{SWAP})&=\frac{d^2-1}{2d^2}(e_u-2g_u+2),
\end{aligned}
\end{equation}
where we have used $E(\text{SWAP})=(d^2-1)/d^2$. Then, we obtain by Eq.~\eqref{eq-operator-entanglement-rewitten},
\begin{equation}
\begin{aligned}
\bra{I}\otimes \bra{S} u^{\otimes 2,2} \ket{I} \otimes \ket{S}&=d^4
\left(1-\frac{d^2-1}{2d^2}(e_u+2g_u) \right), \label{eq:matrix_element_u_ISIS}\\
\bra{I}\otimes \bra{S} u^{\otimes 2,2} \ket{S} \otimes \ket{I}&=d^4
\left(1-\frac{d^2-1}{2d^2}(e_u-2g_u+2) \right).
\end{aligned}
\end{equation}

We represent the operator $W^u$ in the orthonormal basis $\bigl\{
\ket{\overline{0},\overline{0}},\;
\ket{\overline{0},\overline{1}},\;
\ket{\overline{1},\overline{0}},\;
\ket{\overline{1},\overline{1}}
\bigr\}$,
where $\ket{s_1,s_2}=\ket{s_1}\otimes\ket{s_2}$ with $s_i\in\{\overline{0},\overline{1}\}$ and $\ket{\overline{0}}$ and $\ket{\overline{1}}$ are defined in Eq.~\eqref{eq:def_overline_0}.
In this ordered basis, from Eq.~\eqref{eq:matrix_element_u_ISIS}, the matrix form of $W^u$ is 
\begin{align} \label{eq:Wu_matrix_representation}
\begin{pmatrix}
1-\dfrac{d+1}{2(d-1)}\,e_u & 0 & 0 & \dfrac{e_u}{2} \\
0 & 1-g_u & g_u & 0 \\
0 & g_u & 1-g_u & 0 \\
\dfrac{e_u}{2} & 0 & 0 &
1-\dfrac{d-1}{2(d+1)}\,e_u
\end{pmatrix}. 
\end{align}
We give the derivation of $\bra{\overline{0},\overline{0}}W^u\ket{\overline{0},\overline{0}}$ below. The other matrix elements are obtained in a similar way.
Let $U:=u^{\otimes 2,2}$ and use the orthonormal basis vector
\begin{align}
    \ket{\overline{0}}=a\bigl(\ket{I}-\ket{S}\bigr),
\end{align}
where $a={1}/{\sqrt{2d(d-1)}}$ is the scalar coefficient. We then have that
\begin{align}
    \ket{\overline{0},\overline{0}}=a^2\Bigl(\ket{I,I}-\ket{I,S}-\ket{S,I}+\ket{S,S}\Bigr),
\end{align}
and by defining $c_X \in \{1,-1\}$ as the coefficient of $\ket{X}$ in the expansion above for $X=\{II,IS,SI,SS\}$, we obtain that 
\begin{align}
    \bra{\overline{0},\overline{0}}W^u\ket{\overline{0},\overline{0}}
    =a^4\!\!\sum_{X,Y\in\{II,IS,SI,SS\}}\!\! c_X c_Y \bra{X}U\ket{Y}.
\end{align}
Using $U\ket{I,I}=\ket{I,I}$ and $U\ket{S,S}=\ket{S,S}$, we have that 
\begin{align}
    \bra{\overline{0},\overline{0}}W^u\ket{\overline{0},\overline{0}}
    &=a^4\Bigl[
    4d^2(d-1)^2
    +\bra{I,S}U\ket{I,S}+\bra{S,I}U\ket{S,I} \notag\\
    &
    +\bra{I,S}U\ket{S,I}+\bra{S,I}U\ket{I,S}
    -2d^2(d^2-1)
    \Bigr].
\end{align}
Substituting Eq.~\eqref{eq:matrix_element_u_ISIS} to it,
we find that all terms proportional to $g_u$ cancel out, and we obtain 
\begin{align}
    \bra{\overline{0},\overline{0}}W^u\ket{\overline{0},\overline{0}}
    =1-\frac{d+1}{2(d-1)}\,e_u.
\end{align}

\subsection{Solvability condition}
We identify the condition under which a second-moment operator becomes a free-fermionic model.
To this end, we begin by observing that the operator $W^u$ in Eq.~\eqref{eq:Wu_matrix_representation} admits an expansion in the Pauli basis of the form
\begin{equation}
\begin{aligned}
W^u
&=\left(1-\frac{e_u}{4e_{\mathrm{H}}}-\frac{g_u}{4g_{\mathrm{H}}}\right)
\mathbb{I}\mathbb{I}
-\frac{de_u}{2(d^2-1)}
\bigl(Z\mathbb{I}+\mathbb{I}Z\bigr)\\
&-\left(\frac{e_u}{4e_{\mathrm{H}}}-\frac{g_u}{4g_{\mathrm{H}}}\right)
ZZ
+\frac{e_u+2g_u}{4}XX
-\frac{e_u-2g_u}{4}Y Y,
\label{eq:Wu_Pauli_final}
\end{aligned}
\end{equation}
with the qubit-Pauli operators $\{\mathbb{I},X,Y,Z\}$.

We find that if $u$-structured random gates satisfy the condition 
\begin{equation}\label{eq-SM-solvable-condition}
    \frac{e_u}{g_u}=\frac{e_{\mathrm{H}}}{g_{\mathrm{H}}},
\end{equation}
the coefficient of the $ZZ$ interaction in Eq.~\eqref{eq:Wu_Pauli_final} becomes zero, and the operator $W^u$ can be regarded as an interaction term of the $XY$ model. 
The $XY$ model in a one-dimensional system can be mapped to free fermions via Jordan-Wigner transformation \cite{LIEB1961407}, and the Hamiltonian is exactly diagonalizable.
Therefore, we call the condition Eq.~\eqref{eq-SM-solvable-condition} the \textit{solvability condition}.
Under the solvability condition, we treat $e_u$ as a free parameter and take $g_u$ as a function of $e_u$. Clearly, the Haar random case where $e_u=e_H$ and $g_u=g_H$ satisfies the condition.

We now turn to identifying the two-qubit gates that satisfy the condition: 
The two-qubit gates take the 
general form 
$u=\exp{-(i/2)(\theta_x XX+\theta_y YY+\theta_z ZZ)}$, up to single-qubit gates, where $\theta_x$, $\theta_y$, and $\theta_z$ are real numbers \cite{PhysRevA.63.062309}. 
Here, the choice of single-qubit gates is irrelevant to the moment operator, since structured random gates are sandwiched by single-qudit Haar random gates. With this parametrization, the solvability condition on the parameters is
\begin{equation}
    f(\theta_x, \theta_y)+f(\theta_y, \theta_z)+f(\theta_z, \theta_x)=0, \label{eq:solvability_condition_qubit}
\end{equation}
where $f(x_1,x_2)=\sin^2(x_1)[\cos^2(x_2)-{3}/{5}]$. For example, when $\theta_x=\theta_y=\theta_z$, the solvability condition yields  $u=\exp{-i (\theta/2)( XX+YY+ZZ)}$ with $\cos{ \theta}=\pm \sqrt{
{3}/{5}}$.
A derivation of Eq.~\eqref{eq:solvability_condition_qubit} is written in Appendix~\ref{app:solvability_condition_qubit}.
We remark that we can map the moment operators to free-fermion chains under the solvability condition, although the two-qubit gates are in general different from matchgates \cite{10.1145/380752.380785, PhysRevA.65.032325, jozsa2008matchgates}. 
Intuitively, structured random circuits under the solvability condition are free-fermionic ``on average'', while the individual instances are rather chaotic.

\section{Structured local random circuits}
\label{section:Local_SRQC}

In this section, we diagonalize the second-moment operator of a $u$-structured random quantum circuit under the solvability condition.
We also show that its spectral gap increases monotonically with both the entangling power and the gate typicality.
We summarize the results of the section below.

To state the theorem, we define two sets as
\begin{equation} \label{eq:set_wave_number}
\begin{aligned}
        K_0 &\coloneqq  \left\{
        k=\frac{(2m-1)\pi}{n}
        \,\middle|\,
        m=-\frac{n}{2}+1,-\frac{n}{2},\dots,\frac{n}{2}
        \right\}, \\
        K_1 &\coloneqq  \left\{
        k=\frac{2m\pi}{n}
        \,\middle|\,
        m=-\frac{n}{2}+1,-\frac{n}{2},\dots,\frac{n}{2}
        \right\},
\end{aligned}
\end{equation}
where $n$ is the number of qudits. 
The exact spectrum is obtained under the solvability condition.

\begin{thm}[Exact spectrum] \label{thm:exact_spectrum_local}
Let $u$ be a two-qudit unitary gate satisfying the solvability condition~\eqref{eq-SM-solvable-condition}.
Then, the eigenvalues of $M_{\mathrm{L},u}$ are given by 
\begin{equation}
    \begin{aligned}
        E_u(\{n_k\})= 1-\frac{1}{n}\sum_{k \in K_p}\epsilon_kn_k,
    \end{aligned}
\end{equation}
where $p\in\{0,1\}$, $\{n_k\}_{k \in K_p} $ $\in \{0,1\}^{\times n}$ such that $\sum_{k \in K_p} n_k$ is even, and 
\begin{equation}
     \epsilon_k \coloneq  \frac{e_u}{e_{\mathrm{H}}} \left( 1 - \frac{2d}{d^2+1} \cos k \right).
\end{equation}
In particular, the spectral gap of $M_{\mathrm{L},u}$ is
\begin{equation}
    \Delta_{\mathrm{L},u}=\frac{2e_u}{ne_{\mathrm{H}}} \left( 1 - \frac{2d}{d^2+1} \cos \frac{\pi}{n} \right).
\end{equation}
\end{thm}

\noindent 
The proof for Theorem~\ref{thm:exact_spectrum_local} is given in Section~\ref{subsection:exact_diagonalization_local_random_structured}.

Beyond the solvability condition, we obtain that the monotonicity of $u$-structured local random circuits with respect to the positivity and spectral gap as the entangling power and the gate typicality of $u$ increase. 
Specifically, we show that structured local random circuits with large $e_u$ and $g_u$ are more random than the corresponding Haar local random circuits.

\begin{thm}[More global and less local randomness] \label{thm:more_randomness_loca_random}
    Suppose a two-qudit unitary gate $u$ satisfies both $e_u > e_{\mathrm{H}}$ and $g_u > g_{\mathrm{H}}$. Then, $u$-structured local random circuits $\nu_u^{L}$ are more random than the corresponding Haar local random circuits $\nu_\mathrm{H}^{L}$ with respect to both the spectral gap and positivity, in the sense of Definitions~\ref{def-more-randomness-gap} and \ref{def-more-randomness-positivity}.
\end{thm}

\noindent
The proof for the part of positivity is provided in Section~\ref{sec-more-randomness-positivity-LRC}. The proof for the part of the spectral gap combines the results of Theorem~\ref{thm:exact_spectrum_local} and more randomness with respect to positivity, which is given in Section~\ref{sec-more-randomness-gap-LRC}.

\subsection{Exact diagonalization of $XY$ models} \label{section:exact_XY_model}
We briefly review the exact diagonalization of $XY$ models by the Jordan-Wigner transformation \cite{LIEB1961407}.
Readers familiar with free-fermion techniques may skip this subsection and proceed directly to Section~\ref{subsection:exact_diagonalization_local_random_structured}.

Let us consider an $n$-qubit system with an even number $n$. The Hamiltonian $H_{XY}$ of a one-dimensional $XY$ model is given by
\begin{equation}
    -J \sum_{j=1}^{n} \left( \frac{1+\gamma}{2} X_j X_{j+1} + \frac{1-\gamma}{2} Y_j Y_{j+1} \right) - h \sum_{j=1}^n Z_j,
\end{equation}
where we consider the regime $J>0$, $\gamma \in [0, 1]$, and $h$ is a real number. We impose the periodic boundary condition on the system.
We map the spin operators to fermionic operators $c_j$ and $c_j^\dagger$ by the Jordan-Wigner transformation:
\begin{equation} \label{eq:def_JW_transformation}
    c_j = \left( \prod_{l=1}^{j-1} -Z_l \right) \sigma_j^-,
\end{equation}
where the lowering operator is $\sigma_j^- = \frac{1}{2}(X_j - iY_j)$
and the raising operator is  $\sigma_j^+ = (X_j+iY_j)/2$.
The fermionic operators satisfy the relations $\{c_i,c_j^\dagger\}=\delta_{i,j}$.
The periodic boundary condition, namely $\sigma_{n+1}^\pm = \sigma_1^\pm$, implies the following boundary condition for the fermionic operators:
\begin{equation}
    c_{n+1} = -(-1)^N c_1,
    \label{eq:fermion_BC}
\end{equation}
where $N = \sum_{j=1}^n c_j^\dagger c_j$ is the total fermion number operator.

By this mapping, the Hamiltonian $H_{XY}$ is written in terms of fermionic operators as 
\begin{equation} \label{eq:XY_Hamiltonian_real_space}
\begin{aligned}
    H_{XY} = &-\sum_{j=1}^n \Big( J(c_j^\dagger c_{j+1} + c_{j+1}^\dagger c_j) \\
    &+ {J\gamma}(c_j^\dagger c_{j+1}^\dagger + c_{j+1} c_j) 
    + 2h c_j^\dagger c_j- h \Big).
\end{aligned}
\end{equation}
The Hamiltonian commutes with the parity operator $(-1)^N$, and the system decomposes into two parity sectors: the even and odd sectors, where $(-1)^N=1$ and $(-1)^N=-1$, respectively.
When the parity is even (odd), the boundary condition becomes
antiperiodic $c_{n+1} = - c_1$ (periodic $c_{n+1} =  c_1$).

The Hamiltonian can be diagonalized in the momentum space via a Fourier transform:
\begin{equation}
    c_k =\frac{1}{\sqrt{n}} \sum_{j=1}^n e^{-ikj} c_j.
\end{equation}
We use the same notation of $c$-fermions for the real and momentum spaces, as we believe that there will be no confusion.
In each parity sector, the values that $k$ can take depend on the parity so that the fermionic operators satisfy the boundary conditions. 
Concretely, the sets of the wave numbers $k$, denoted by $ K_p$ for $p=0,1$ are defined by Eq.~\eqref{eq:set_wave_number}.
Due to the assumption that $n$ is even, the values of $m$ are integers in Eq.~\eqref{eq:set_wave_number}.
Here $p=0$ ($p=1$) refers to the even (odd) sector.
With these definitions, the inverse Fourier transform is given by
\begin{equation} \label{eq:Fourier_transform_inverse}
    c_j=\frac{1}{\sqrt{n}}\sum_{k \in K_p} e^{ikj} c_k,
\end{equation}
where $K_p$ depends on the parity sector on which $c_j$ acts. 
We note that the total fermion number operator transforms as $N = \sum_{j=1}^n c_j^\dagger c_j = \sum_{k \in K_p} c_k^\dagger c_k$.
We can then write the Hamiltonian Eq.~\eqref{eq:XY_Hamiltonian_real_space} in the momentum space as
\begin{equation}
    H_{XY} = \sum_{k \in K_p} \left[ \xi_k c_k^\dagger c_k - \frac{i}{2} \Delta_k (c_k^\dagger c_{-k}^\dagger - c_{-k} c_k) \right] + nh,
\end{equation}
where $\xi_k = - 2(J \cos k + h)$ and $\Delta_k = 2J \gamma \sin k$.
The Hamiltonian has the \emph{Bogoliubov-de Gennes} (BdG) form \cite{de2018superconductivity}, 
\begin{equation} \label{eq:Hamiltonian_XY_BdG}
    \frac{1}{2} \sum_{k \in K_p}\Psi_k^{\dagger}\begin{pmatrix}
    \xi_k & -i\Delta_k \\
    i\Delta_k & -\xi_k
    \end{pmatrix}\Psi_k,
\end{equation}
where $\Psi_k =(c_k,c_{-k}^\dagger )^{\trans}$ is the Nambu spinor. The Bogoliubov transformation to quasiparticles $\Phi_k = (\eta_k, \eta_{-k}^\dagger)^{\trans}$ is defined by $\Phi_k = U_k \Psi_k$, with the unitary matrix
\begin{equation}\label{eq:Bogoliubov_transofrmation_unitary}
    U_k = \begin{pmatrix}
    \cos \frac{\theta_k}{2} & -i \sin \frac{\theta_k}{2} \\
    -i \sin \frac{\theta_k}{2} & \cos \frac{\theta_k}{2}
    \end{pmatrix},
\end{equation}
where  $\cos\theta_k = \xi_k / \epsilon_k$, $\sin\theta_k = \Delta_k / \epsilon_k$, and $ \epsilon_k = \sqrt{\xi_k^2 + \Delta_k^2}$.
This transformation diagonalizes the BdG form, as
\begin{equation}
\begin{aligned}
     H_{XY}= \sum_{k \in K_p} \epsilon_k \left( \eta_k^\dagger \eta_k - \frac{1}{2} \right),
    \label{eq:H_XY_diagonal_quasi_particle}
\end{aligned}
\end{equation}
with the single-particle energy 
\begin{equation} \label{eq:single_particle_energy}
    \epsilon_k = 2\sqrt{(J \cos k + h)^2 + (J \gamma \sin k)^2}.
\end{equation}

In $p$-parity sector, the ground state vector of the Hamiltonian is obtained by 
\begin{equation} \label{eq:ground_state_general}
    \ket{\mathrm{GS}}_p = \left( c_0^\dagger \right)^{p} \prod_{k \in K_p: 0<k<\pi} \left( \cos \frac{\theta_k}{2} + i \sin \frac{\theta_k}{2} c_k^\dagger c_{-k}^\dagger \right) \ket{\mathrm{vac}}_c,
\end{equation}
where $\ket{\mathrm{vac}}_c$ is the vacuum of the fermions $c_k$, namely, for any $k\in K_p$, $c_k\ket{\mathrm{vac}}_c=0$.
Here, for the purpose of the subsequent discussion, we have assumed that $J>\abs{h}$ and hence $\xi_0 < 0$, so that the zero-momentum mode must be occupied in the ground state, yielding  $c_0^\dagger$ in the odd sector ($p=1$) in Eq.~\eqref{eq:ground_state_general}. 
The occupation of the $k=0$ mode guarantees that the ground state in the odd sector contains an odd number of fermionic operators $c_k$.
The ground states satisfies $\eta_k\ket{\mathrm{GS}}_p=0$ for any $k\in K_p$, and the corresponding eigenenergy is 
\begin{equation} \label{eq:XY_model_ground_energy}
    E_{\mathrm{GS}} = -\frac{1}{2}\sum_{k \in K_p}\epsilon_k.
\end{equation}
Moreover, for occupation numbers $\{n_k\}_{k \in K_p} \in \{0,1\}^{\times n}$ such that $\sum_{k \in K_p}n_k$ is even, the eigenvectors of $H_{XY}$ are 
\begin{equation} \label{eq:XY_model_eigenvectors}
    \ket{\{n_k\}} = \prod_{k \in K_p} (\eta_k^\dagger)^{n_k} \ket{\mathrm{GS}}_p,
\end{equation}
and the corresponding eigenvalues are 
\begin{equation} \label{eq:XY_energy_n_k}
    \begin{aligned}
        E_{\{n_k\}} = E_{\mathrm{GS}} + \sum_{k \in K_p} \epsilon_k n_k.
    \end{aligned}
\end{equation}
Requiring $\sum_{k \in K_p} n_k$ to be even guarantees that the parity of $ \ket{\{n_k\}}$ coincides with that of the ground state.

\subsection{Exact diagonalization of structured local random circuits}
\label{subsection:exact_diagonalization_local_random_structured}
We apply the exact diagonalization of the $XY$ models to the second-moment operators of structured local random circuits under the solvability condition.
As we discussed in Section~\ref{subsec:secomd_moment_structured_gates}, the second-moment operator of a $u$-structured local random circuit is 
\begin{equation}
     M_{\mathrm{L},u} = \frac{1}{n} \sum_{i=1}^n W^u_{i,i+1},
\end{equation}
where $W^u_{i,i+1}$ is given by Eq.~\eqref{eq:Wu_matrix_representation}. From Eq.~\eqref{eq:Wu_Pauli_final}, we have that
\begin{align}
    M_{L,u} =I-\frac{1}{n}H_{u},
\end{align}
where we write the Hamiltonian $H_u$ as
\begin{align} \label{eq:H_u_Hamiltonian_general}
    \begin{aligned}
    &H_{u}= \, n \! \left(\frac{e_u}{4e_{\mathrm{H}}}+\frac{g_u}{4g_{\mathrm{H}}}\right)I
    -\sum_{i=1}^n 
    \Biggl( \frac{e_u+2g_u}{4}X_iX_{i+1}\\
    &-\frac{e_u-2g_u}{4}\,Y_iY_{i+1}
     -\left(\frac{e_u}{4e_{\mathrm{H}}}
    -\frac{g_u}{4g_{\mathrm{H}}}\right) \! Z_i Z_{i+1}-\frac{e_u d}{d^2-1}Z_i \Biggl)
    \end{aligned}
\end{align}
When the solvability condition holds, it becomes
\begin{equation} \label{eq:H_u_XY_model}
\begin{aligned}
    H_u=\frac{n e_u}{2e_{\mathrm{H}}}I
    - \frac{e_u}{2(d^2-1)} \sum_{i=1}^n (
     d^2 X_i X_{i+1} 
        +  Y_i Y_{i+1}
        -2d Z_i).
\end{aligned}
\end{equation}
Up to an additive constant, we identify this as the Hamiltonian of the $XY$ model with the parameters
\begin{equation}
\begin{aligned}
J &= \frac{e_u}{2e_{\mathrm{H}}} \\
\gamma &= e_{\mathrm{H}} \\
h &= - \frac{e_u d}{d^2-1}.
\end{aligned}
\end{equation}
The parameters in the BdG form Eq.~\eqref{eq:Hamiltonian_XY_BdG} are given by
\begin{equation}
\begin{aligned}
    \xi_k 
    &= -\frac{e_u}{e_{\mathrm{H}}} \left( \cos k - \frac{2d}{d^2+1} \right),\\
    \Delta_k &= e_u \sin k.
\end{aligned}
\end{equation}
Then the Bogoliubov transformation turns out to be
\begin{equation} \label{eq:bogoliubov_eta_k}
\begin{aligned}
    \begin{pmatrix}
        \eta_k \\
        \eta_{-k}^\dagger
    \end{pmatrix}
    &=
    \begin{pmatrix}
        \cos \frac{\theta_k}{2} & -i \sin \frac{\theta_k}{2} \\
        -i \sin \frac{\theta_k}{2} & \cos \frac{\theta_k}{2}
    \end{pmatrix}
    \begin{pmatrix}
        c_k \\
        c_{-k}^\dagger
    \end{pmatrix}\\
     &= \frac{1}{\sqrt{2\epsilon_k}}
    \begin{pmatrix}
        \sqrt{\epsilon_k+\xi_k} & -i \sqrt{\epsilon_k-\xi_k} \\
        -i \sqrt{\epsilon_k-\xi_k} & \sqrt{\epsilon_k+\xi_k}
    \end{pmatrix}
    \begin{pmatrix}
        c_k \\
        c_{-k}^\dagger
    \end{pmatrix}.
\end{aligned}
\end{equation}
We also obtain the single-particle energy from Eq.~\eqref{eq:single_particle_energy},
\begin{equation} \label{eq:single_particle_energy_structured_local}
\begin{aligned}
    \epsilon_k &= 2\sqrt{ h^2 + 2hJ \cos k + J^2 \cos^2 k + J^2 \gamma^2 \sin^2 k}\\
    &= \frac{e_u}{e_{\mathrm{H}}} \left( 1 - \frac{2d}{d^2+1} \cos k \right).
\end{aligned}
\end{equation}

In the basis of $\{\ket{\overline{0}},\ket{\overline{1}}\}^{\otimes n}$, the eigenvectors of $Z$ are $\ket{\overline{0}}$ and $\ket{\overline{1}}$ with the eigenvalues $1$ and $-1$, respectively, and the vacuum is given by $\ket{\mathrm{vac}}_c=\ket{\overline{1}}^{\otimes n}$. We note that in this basis, $c$-fermion is occupied for $\ket{\overline{0}}$ and not occupied for $\ket{\overline{1}}$.
We derive the ground states $\ket{\mathrm{GS}}_p$. Because we know that the second-moment operator $M_{\mathrm{L},u}$ has two eigenvectors with eigenvalue $1$, namely $\ket{I}^{\otimes n}$ and $\ket{S}^{\otimes n}$, the ground states are superpositions of them,
\begin{align}
    \ket{\mathrm{GS}}_p=a_{0p}\ket{I}^{\otimes n}+a_{1p}\ket{S}^{\otimes n}.
\end{align}
We expand as $\ket{I}=v_0\ket{\overline{0}}+v_1\ket{\overline{1}}$ and $\ket{S}=-v_0\ket{\overline{0}}+v_1\ket{\overline{1}}$, where $v_0=\sqrt{d(d-1)/2}$ and $v_1=\sqrt{d(d+1)/2}$. It implies that 
\begin{equation}
\begin{aligned}
    \ket{I}^{\otimes n}&=\sum_{i_1,i_2,...,i_n \in \{0,1\}}\prod_{j=1}^nv_{i_j}\ket{\overline{i_1}, \overline{i_2} ,\cdots ,\overline{i_n}},\\
    \ket{S}^{\otimes n}&=\sum_{i_1,i_2,...,i_n \in \{0,1\}}\prod_{j=1}^n(-1)^{i_j}v_{i_j}\ket{\overline{i_1} , \overline{i_2} ,\cdots,  \overline{i_n}},
\end{aligned}
\end{equation}
and in the vector $\ket{S}^{\otimes n}$, the coefficient of the basis state is positive if and only if the number of occurrences of $\overline{1}$ is even. Since $n$ is even, the parity of the number of $\overline{1}$ is equal to that of $\overline{0}$.  
Therefore, the ground state vectors in even and odd sectors are constructed as
\begin{equation}
\begin{aligned}
    \ket{\mathrm{GS}}_0 &=\frac{1}{\sqrt{2d^n(d^n+1)}}\left(\ket{I}^{\otimes n}+\ket{S}^{\otimes n}\right), \\
    \ket{\mathrm{GS}}_1 &=\frac{1}{\sqrt{2d^n(d^n - 1)}}\left(\ket{I}^{\otimes n}-\ket{S}^{\otimes n}\right),
\end{aligned}
\end{equation}
where we have normalized the states by using $\braket{\sigma}{\tau}=d+\delta_{\sigma,\tau}(d^2-d)$, for $\sigma, \tau \in \{I,S\}$.

The other eigenvectors are obtained by adding $\eta$-fermions to the ground states. 
Explicitly, for parity $p$ and occupation numbers $\{n_k\}_{k \in K_p} \in \{0,1\}^{\times n}$ satisfying the even-parity condition $\sum_{k \in K_p}n_k \in 2\mathbb{Z}$, the eigenvectors  are 
\begin{equation} \label{eq:structured_local_eigenvectors}
    \ket{\{n_k\}} = \prod_{k \in K_p} (\eta_k^\dagger)^{n_k} \ket{\mathrm{GS}}_p,
\end{equation}
where $\eta_k$ is given by Eq.~\eqref{eq:bogoliubov_eta_k}. 
The corresponding eigenvalues of $H_u$ are then given by Eq.~\eqref{eq:XY_energy_n_k}: for parity $p$ and $\{n_k\}_{k \in K_p} \in \{0,1\}^{\times n}$ such that $\sum_{k \in K_p} n_k \in 2 \mathbb{Z}$, 
\begin{equation}
    \begin{aligned}
        E_u(\{n_k\})&=\frac{ne_u}{2e_{\mathrm{H}}}+\sum_{k \in K_p}\epsilon_k \left(n_k -\frac{1}{2} \right)  \\
        &= \sum_{k \in K_p}\epsilon_kn_k.
    \end{aligned}
\end{equation}
We derive the first excited energy. Since $\epsilon_k$ is monotonically increasing with $|k|$, for $0 \leq \abs{k} \leq \pi$, the first excited state in even (odd) sector is obtained by adding two quasi-particles with $k=\pm \pi/n \in K_0$ ($k=0, 2\pi/n \in K_1$) to the ground state. It implies that the first excited energies in the even sector $E^{(1)}_{u,\mathrm{even}}$ and in the odd sector $E^{(1)}_{u,\mathrm{odd}}$ are given by
\begin{equation}
\begin{aligned}
    E^{(1)}_{u,\mathrm{even}}&=\epsilon_{\frac{\pi}{n}} + \epsilon_{ -\frac{\pi}{n}}\\
    &=\frac{2e_u}{e_{\mathrm{H}}} \left( 1 - \frac{2d}{d^2+1} \cos \frac{\pi}{n} \right),
\end{aligned}
\end{equation}
\begin{equation}
    \begin{aligned}
        E^{(1)}_{u,\mathrm{odd}}&=\epsilon_{0} + \epsilon_{ \frac{2\pi}{n}}\\
        &= \frac{e_u}{e_{\mathrm{H}}} \left[ \left(1 - \frac{2d}{d^2+1}\right) + \left(1 - \frac{2d}{d^2+1} \cos \frac{2\pi}{n}\right) \right],
    \end{aligned}
\end{equation}
where we have used Eq.~\eqref{eq:single_particle_energy_structured_local}.
Because $\cos{k}$ is a strictly concave function, we have that $E^{(1)}_{u,\mathrm{odd}} >E^{(1)}_{u,\mathrm{even}}$ for a finite $n$. The first excited energy is then given by $E^{(1)}_{u,\mathrm{even}}$.
On the other hand, the largest eigenvalue is obtained by filling all $\eta$-fermions, and it is
 \begin{align}
     &\sum_{k} \frac{e_u}{e_{\mathrm{H}}} 
    \left( 1-\frac{2d}{d^2+1}\cos k \right)=\frac{ne_u}{e_{\mathrm{H}}}.
\label{eq:evalmin_lower_bound}
 \end{align}
From $M_{L,u} =\mathbb{I}-\frac{1}{n}H_{u}$, the corresponding eigenvalue of $M_{L,u}$ is negative when $e_u>e_{\mathrm{H}}$, and the absolute value of $1-e_u/e_{\textrm{H}}$ for $e_{\mathrm{H}}<e_u \leq 1$ is less than $1-(1/n)E^{(1)}_{u,\mathrm{even}}$ for $n \geq 2$.
Therefore, we obtain the spectral gap $\Delta_{L,u}$ of the second-moment operator $M_{L,u}$ as
\begin{equation} \label{eq:spectral_gap_local_random_proof}
    \Delta_{\mathrm{L},u}=\frac{2e_u}{ne_{\mathrm{H}}} \left( 1 - \frac{2d}{d^2+1} \cos \frac{\pi}{n} \right).
\end{equation}

\subsection{Beyond free fermions}
\label{sec-more-randomness-positivity-LRC}
We analyze the second-moment operator beyond the regime satisfying the solvability condition. The main goal in this subsection is to prove that $u$-structured random circuits are more random with respect to positivity than Haar local random circuits if $e_u>e_\mathrm{H}$ and $g_u>g_\mathrm{H}$.

We begin by observing that more randomness with respect to positivity is trivially true under the solvability condition. 
From Eq.~\eqref{eq:H_u_XY_model}, the operator $H_u$ is proportional to $\frac{e_u}{e_{\mathrm{H}}}$, which implies that the eigenvalues of $H_u$ monotonically increase in $e_u$. Moreover, it is clear from Eq.~\eqref{eq:bogoliubov_eta_k} that the eigenvectors of $H_u$ do not depend on $e_u$. Thus by $M_{\mathrm{L},u}=I-(1/n)H_u$, we have that, when $e_u>e_{\mathrm{H}}$,
\begin{align}
    \left. M_{\mathrm{L},u} \right|_{V_1^\perp} < \left. M_{\mathrm{L},\mathrm{H}} \right|_{V_1^\perp}, \label{eq-more-randomness-solvable-condition}
\end{align}
where we have used the fact that $V_1=\mathrm{span} \{ \ket{I}^{\otimes n},\ket{S}^{\otimes n}\}$ is the ground eigenspace of $H_u$.

We now consider the general case, in which $e_u$ and $g_u$ do not necessarily satisfy the solvability condition. 
From Eq.~\eqref{eq:H_u_Hamiltonian_general}, we have that  $H_u = ({e_u}/{e_{\mathrm{H}}}) H_e + ({g_u}/{g_{\mathrm{H}}}) H_g$, where $H_a=\sum_{i=1}^nh_{a,i}$, for $a=e,g$, with  
\begin{equation}\label{eq:he_hg_corrected}
\begin{aligned}
    h_{e,i} & = \frac{1}{4}I + \frac{1}{4}Z_i Z_{i+1} + \frac{d}{2(d^2+1)}(Z_i + Z_{i+1}) \\
   & \qquad  - \frac{e_{\mathrm{H}}}{4}(X_i X_{i+1} - Y_i Y_{i+1}),\\
    h_{g,i} & = \frac{1}{4}I - \frac{1}{4}Z_i Z_{i+1} - \frac{1}{4}(X_i X_{i+1} + Y_i Y_{i+1}).
\end{aligned}
\end{equation}
In the basis $\{
\ket{\bar{0},\bar{0}},
\ket{\bar{0},\bar{1}},
\ket{\bar{1},\bar{0}},
\ket{\bar{1},\bar{1}}
\}$, the matrix representations are given by 
\begin{equation}\label{eq:matrix_h_e}
    h_e = 
    \begin{pmatrix}
        \frac{(d+1)^2}{2(d^2+1)} & 0 & 0 & -\frac{e_{\mathrm{H}}}{2} \\
        0 & 0 & 0 & 0 \\
        0 & 0 & 0 & 0 \\
        -\frac{e_{\mathrm{H}}}{2} & 0 & 0 & \frac{(d-1)^2}{2(d^2+1)}
    \end{pmatrix},
\end{equation}
\begin{equation}\label{eq:matrix_h_g}
    h_g = 
    \begin{pmatrix}
        0 & 0 & 0 & 0 \\
        0 & \frac{1}{2} & -\frac{1}{2} & 0 \\
        0 & -\frac{1}{2} & \frac{1}{2} & 0 \\
        0 & 0 & 0 & 0
    \end{pmatrix}.
\end{equation}
The eigenspectrum of $h_e$ is $\{1, 0, 0, 0\}$, and hence $h_e$ is positive semidefinite. The matrix $h_g$ is also positive semidefinite and the eigenspectrum of $h_e$ is $\{1, 0, 0, 0\}$.
The Hamiltonian $H_u$ is a sum of $h_e$ and $h_g$, and thus it is also positive semidefinite.

From the above discussion, the second-moment operator $M_{\mathrm{L},u}=\mathbb{I}-(1/n)H_u$ monotonically decreases 
in $e_u$ and $g_u$, in the sense that 
\begin{align}
    & M_{\mathrm{L},u}(e_u+e',g_u+g') \leq M_{\mathrm{L},u}(e_u,g_u), \label{SM-eq-operator-ineq-moment-operator}
\end{align}
 for $e',g' \geq 0$, where we explicitly write $M_{\mathrm{L},u}$ as a function of $e_u$ and $g_u$.
It implies that, if $e_u>e_{\mathrm{H}}$ and $g_u > g_\mathrm{H}$, we have
\begin{align}
    \left. M_{\mathrm{L},u}\right|_{V_1^\perp} < \left.  M_{\mathrm{L},\mathrm{H}}\right|_{V_1^\perp}, \label{SM-eq-more-random-positive}
\end{align}
where we have used that $\Delta M_{\mathrm{L},u}\coloneq M_{\mathrm{L},u}(e_u,g_u)-M_{\mathrm{L},u}(e_u+e',g_u+g')$ is strictly positive  for $e',g' > 0$, because the kernel of $H_u$ is exactly $V_1$ when $e_u>0$.

We remark that the above proof is independent of the circuit architecture, and therefore the operator inequality applies to structured local random circuits in arbitrary spatial dimensions and with arbitrary connectivity.
In addition, the moment operators $M_{\mathrm{L},u}$ without the solvability condition are equivalent to frustration-free $XYZ$ chains in a magnetic field, which are mapped to interacting Kitaev chains \cite{PhysRevB.92.115137, PhysRevB.95.195140, vardhan2024entanglement}.

\subsection{Monotonicity of gap in structured local random circuits}
\label{sec-more-randomness-gap-LRC}
We now prove that $u$-structured random circuits exhibit a larger spectral gap than local Haar random circuits when $e_u>e_\mathrm{H}$ and $g_u>g_\mathrm{H}$.
Given the monotonic operator inequality in Eq.~\eqref{SM-eq-more-random-positive}, Weyl's monotonicity theorem \cite{bhatia2013matrix} guarantees that the eigenvalues $\{\lambda_i\}$ of $\left. M_{\mathrm{L},u} \right|_{V_1^\perp}$ also monotonically decrease as $e_u$ and $g_u$ increase. 
The spectral gap of $M_{\mathrm{L},u}(e_u,g_u)$ then increases monotonically with $e_u$ and $g_u$, when the eigenvalue with the second-largest absolute value is strictly positive.
In the following, we prove that the second-largest eigenvalue of $M_{\mathrm{L},u}$ in absolute value is positive whenever $e_u>e_\mathrm{H}$ and $g_u>g_\mathrm{H}$. 

We write the maximum and minimum eigenvalues of $\left. M_{\mathrm{L},u} \right|_{V_1^\perp}$ by $\lambda_{\mathrm{max}}(e_u,g_u)$ and $\lambda_{\mathrm{min}}(e_u,g_u)$, respectively. 
We consider the regime $e_u>e_\mathrm{H}$ and $g_u>g_\mathrm{H}$, and we aim to show that
\begin{equation} \label{eq:second-largest_eigen_value_positive}
    |\lambda_{\mathrm{min}}(e_u,g_u)| \leq \lambda_{\mathrm{max}}(e_u,g_u),
\end{equation}
which implies that the second-largest eigenvalue in absolute value is positive.
We can focus on the case where $\lambda_{\mathrm{min}}(e_u,g_u)$ is negative, since otherwise it is trivial that $|\lambda_{\mathrm{min}}(e_u,g_u)| = \lambda_{\mathrm{min}}(e_u,g_u) \leq \lambda_{\mathrm{max}}(e_u,g_u)$. 
To prove Eq.~\eqref{eq:second-largest_eigen_value_positive}, we derive a lower bound on the eigenvalues of $M_{\mathrm{L},u}$ by using the exact results under the solvability condition. 
We define the parameter 
\begin{equation}
   c_{u} \coloneq \max \left\{ \frac{e_u}{e_{\mathrm{H}}}, \frac{g_u}{g_{\mathrm{H}}} \right\}.
\end{equation}
Then, from Eq.~\eqref{SM-eq-more-random-positive}, it implies that 
\begin{equation}
    M_{\mathrm{L},u}(e_{\mathrm{H}}c_{u},g_{\mathrm{H}}c_{u}) \leq M_{\mathrm{L},u}(e_u,g_u). \label{eq-SM-Ml-operatorineq-min} 
\end{equation}
From this operator inequality, the eigenvalues are bounded as
\begin{equation}
\begin{aligned}
    \lambda_{\mathrm{max}}(e_{\mathrm{H}}c_{u},g_{\mathrm{H}}c_{u}) &\leq \lambda_{\mathrm{max}}(e_u,g_u),\\
    \lambda_{\mathrm{min}}(e_{\mathrm{H}}c_{u},g_{\mathrm{H}}c_{u}) &\leq \lambda_{\mathrm{min}}(e_u,g_u).
\end{aligned}
\end{equation}
From Eq.~\eqref{eq:spectral_gap_local_random_proof}, the maximum eigenvalue at the solvable point is given by
\begin{equation}
    \lambda_{\mathrm{max}}(e_{\mathrm{H}}c_{u},g_{\mathrm{H}}c_{u}) = 1-\frac{2c_u}{n} \left( 1-\frac{2d}{d^2+1}\cos\frac{\pi}{n} \right). \label{eq-lmax_lower} 
\end{equation}
Moreover, from Eq.~\eqref{eq:evalmin_lower_bound}, the minimum eigenvalue is bounded by
\begin{equation}
     1-c_u \leq \lambda_{\mathrm{min}}(e_{\mathrm{H}}c_{u},g_{\mathrm{H}}c_{u}). \label{eq-lmin_lower} 
\end{equation}

We derive upper bounds on $c_u$. Since the feasible values of $e_u$ are $0 \leq e_u \leq \frac{2}{3}$ ($d=2$) and $0 \leq e_u \leq 1$ ($d \geq 3$), we have that
\begin{equation}
    \begin{aligned}
        \frac{e_u}{e_{\mathrm{H}}} &\leq \frac{10}{9}  &&(d=2),\\
        \frac{e_u}{e_{\mathrm{H}}} &\leq \frac{d^2+1}{d^2-1} &&(d \geq 3).
    \end{aligned}
\end{equation}
Furthermore, under the assumption $e_u \geq e_{\mathrm{H}}$, Eq.~\eqref{eq:eu-gu-upper-boundary} implies that  $\frac{g_u}{g_{\mathrm{H}}} \leq 2-e_{\mathrm{H}}$. This yields that 
\begin{equation}
    \begin{aligned}
        \frac{g_u}{g_{\mathrm{H}}} &\leq \frac{7}{5}  &&(d=2),\\
        \frac{g_u}{g_{\mathrm{H}}} &\leq 2-\frac{d^2-1}{d^2+1} &&(d \geq 3).
    \end{aligned}
\end{equation}
Combining them, we upper bound $c_u$ as
\begin{equation}
\begin{aligned}
    c_u &\leq \frac{7}{5}  &&(d=2), \label{eq:c_u-d=2} \\
    c_u &\leq \frac{d^2+1}{d^2-1}  &&(d \geq 3).
\end{aligned}
\end{equation}

We use Eqs.~\eqref{eq-lmax_lower} and \eqref{eq-lmin_lower} to have the difference
\begin{equation}
\begin{aligned}
    \lambda_{\mathrm{max}}-|\lambda_{\mathrm{min}}| &= \lambda_{\mathrm{max}} + \lambda_{\mathrm{min}} \\
    &\geq 2 - c_u\left[ \frac{2}{n}\left( 1-\frac{2d}{d^2+1}\cos\frac{\pi}{n} \right) + 1 \right]. \label{eq:diff_lambda_max_min}
\end{aligned}
\end{equation}
Since ${2}/{n}$ and $1-({2d}\cos({\pi}/{n}))/({d^2+1})$ are both positive and monotonically decreasing with $n$, their product is also monotonically decreasing. 
The right-hand side of Eq.~\eqref{eq:diff_lambda_max_min} then increases monotonically with $n$. By calculating it at the most stringent bound for $n=3$, we obtain
\begin{equation}
    \lambda_{\mathrm{max}}-|\lambda_{\mathrm{min}}| \geq 2 - c_u \left[ \frac{2(d^2-d+1)}{3(d^2+1)} + 1 \right]. \label{eq:diff_lambda_max_min2}
\end{equation}
For $d=2$, substituting $c_u \leq \frac{7}{5}$ into Eq.~\eqref{eq:diff_lambda_max_min2} yields
\begin{equation}
    \lambda_{\mathrm{max}}-|\lambda_{\mathrm{min}}| \geq \frac{1}{25} > 0.
\end{equation}
For $d \geq 3$, substituting $c_u \leq \frac{d^2+1}{d^2-1}$ into Eq.~\eqref{eq:diff_lambda_max_min2}, we obtain
\begin{equation}
    \lambda_{\mathrm{max}}-|\lambda_{\mathrm{min}}| \geq \frac{1}{3} + \frac{2}{3}\frac{d-2}{d^2-1} - \frac{2}{d^2-1} > 0.
\end{equation}
Therefore, we prove Eq.~\eqref{eq:second-largest_eigen_value_positive} for all $n \geq 3$ and $d \geq 2$.

\section{Structured brick-wall random circuits}
\label{section:brick-wall_SRQC}
We extend the analysis of the previous section to the brick-wall architecture and exactly diagonalize the second-moment operator of structured brick-wall random circuits under the solvability condition.

We summarize the results obtained in this section.
We obtain all the eigenvalues of the second-moment operator under the solvability condition~\eqref{eq-SM-solvable-condition} as follows.

\begin{thm}[Eigenvalues of $M_{\mathrm{B},u}$] \label{thm:exact_spectrum_brick}
Let $u$ be a two-qudit unitary gate satisfying the solvability condition.
Then, the eigenvalues of $M_{\mathrm{B},u}$ are given by 
\begin{equation} \label{eq:eigenvalues.M_Bu_main_thm}
    \prod_{k \in K_p} (\lambda_k)^{n_k}
\end{equation}
where $p \in \{0,1\}$, $\{n_{k}\}_{k\in K_p}$, for $n_{k} \in \{0,1\}$,  satisfies that $\sum_{k \in K_p}n_{k}$ is even, and 
\begin{equation} \label{eq:eigenvalues_lambda_main_thm}
    \lambda_k=\left(
a_k\widetilde e_u
+
\sqrt{
a_k^2\widetilde e_u^2+1-\widetilde e_u
}
\right)^2,
\end{equation}
where $ \widetilde e_u \coloneqq {e_u}/{e_{\mathrm{H}}}$ and $ a_k \coloneqq  ({d}/({d^2+1}))\cos{k}$.
\end{thm}
\noindent
A proof for Theorem~\ref{thm:exact_spectrum_brick} is given in Section~\ref{sec:formula_evalues_brick_wall}.

As a corollary, we obtain the following formula for the exact spectral gap.

\begin{cor}[Spectral gap] \label{cor:spectral_gap_brick-wall}
    The spectral gap of $M_{\mathrm{B},u}$ under the solvability condition is
\begin{equation}
    \Delta_{\mathrm{B},u}=1-\max\left( | \lambda_{\frac{\pi}{n}}|^2, |\lambda_{0}\lambda_{\frac{2\pi}{n}}| \right),
\end{equation}
where $\lambda_k$ is defined in Eq.~\eqref{eq:eigenvalues_lambda_main}. Moreover, when $\lambda_{\frac{2 \pi}{n}}$ is a real number, we have 
\begin{align}
    \Delta_{\mathrm{B},u}=1-\left( \lambda_{\frac{\pi}{n}} \right)^2,
\end{align}
 and when $\lambda_{\frac{ \pi}{n}}$ is not a real number, we have 
 \begin{align}
     \Delta_{\mathrm{B},u}=1-\lambda_{0}\left( \frac{e_u}{e_{\mathrm{H}}} -1 \right).
 \end{align}
\end{cor}
\noindent
A proof for Corollary~\ref{cor:spectral_gap_brick-wall} is given in Section~\ref{sec:spectral_gap_brick}.

Based on the formula, we show the monotonicity of the spectral gap.

\begin{thm}[Monotonicity of the spectral gap] \label{more_randomness_brick_wall}
    Suppose a two-qudit unitary gate $u$ satisfies the solvability condition. Then, $u$-structured brick-wall random circuits $\nu_u^{B}$ are more random than the corresponding Haar brick-wall random circuits $\nu_\mathrm{H}^{B}$ with respect to the spectral gap if and only if $e_u >e_{\mathrm{H}}$.
\end{thm}
\noindent
A proof for Theorem~\ref{more_randomness_brick_wall} is also given in Section~\ref{sec:spectral_gap_brick}.

\subsection{Fermion mapping of brick-wall circuits}
We map the second-moment operators of brick-wall random circuits to fermionic chains via the Jordan-Wigner transformation, as with the previous section. 
The moment operator that we consider is
\begin{equation}
    M_{\mathrm{B},u} =\left(\, \bigotimes_{i \equiv 1 \ (\mathrm{mod} \ 2)}  W^u_{i,i+1}\right)
    \left(\, \bigotimes_{i \equiv0 \ (\mathrm{mod} \ 2)}  W^u_{i,i+1}\right).
\end{equation}
Here, without imposing the solvability condition, we have 
\begin{equation}
\begin{aligned}
W^u_{i,i+1} = I - \frac{e_u}{e_{\mathrm{H}}} h_{e,i} - \frac{g_u}{g_{\mathrm{H}}} h_{g,i},
\end{aligned}
\end{equation}
where the operators $h_{e,i}$ and $h_{g,i}$ are defined in Eq.~\eqref{eq:he_hg_corrected}.

We map the moment operator to a fermionic model. 
To this end, we derive the logarithm $\log W^u_{i,i+1}$ as 
\begin{equation}
\begin{aligned}
\log\left(W^u_{i,i+1}\right) = \log\!\left( 1- \frac{e_u}{e_{\mathrm{H}}} \right) h_{e,i} +\log\!\left( 1- \frac{g_u}{g_{\mathrm{H}}} \right) h_{g,i},
\end{aligned}
\end{equation}
where we have used the fact that $\{I-h_{e,i}-h_{g,i}, h_{e,i} ,h_{g,i} \}$ are orthogonal projectors for each $i$. 
Throughout this work, $\log$ denotes the natural logarithm, namely, $\log \equiv \log_e$.
Here, to take the logarithm, we have assumed that $e_u,g_u>0$, $e_u \neq e_{\mathrm{H}}$, and $g_u \neq g_{\mathrm{H}}$. For a negative number $z$, we take the value $\log(z)=\log(|z|)+i \pi$.  
We note that $M_{B,u}$ is continuous at $e_u=e_{\mathrm{H}}$, and thus the eigenvalues and eigenvectors of $M_{B,\mathrm{H}}$ are given by substituting $e_{\mathrm{H}}$ for $e_u$. 
We now impose the solvability condition~\eqref{eq-SM-solvable-condition}, yielding
\begin{equation} \label{eq:log_Wu}
\begin{aligned}
    \log W^u_{i,i+1} &= \frac{1}{2}\log\!\left(1 - \frac{e_u}{e_{\mathrm{H}}}\right) 
    \biggl( \mathbb{I} -  \frac{d^2}{d^2+1} X_i X_{i+1} \\
    & \quad - \frac{1}{d^2+1} Y_i Y_{i+1} + \frac{d}{d^2+1} (Z_i+Z_{i+1}) \biggr).
\end{aligned}
\end{equation}
We introduce the transfer matrices of the even and odd layers as $M_{\mathrm{B},u}=T_1T_0$, where we define
\begin{equation}
    \begin{aligned}
        T_a \coloneq &\bigotimes_{i \equiv a \ (\mathrm{mod} \ 2)}  W^u_{i,i+1},
    \end{aligned}
\end{equation}
for $a\in \{0,1\}$ and  $i \in \{0,1,...,n\}$.
Since the local operators $W^u_{i,i+1}$ commute mutually, the logarithm of the tensor product becomes the sum 
\begin{equation}
\begin{aligned}
    \log T_a &= \sum_{i \equiv a \ (\mathrm{mod} \ 2)}  \log W^u_{i,i+1}
\end{aligned}
\end{equation}
of the local logarithms.
To diagonalize the moment operator, we express $\log T_a$, for $a=0,1$, in terms of a Nambu spinor. 
We consider the set of wave numbers $K^+_p \coloneq \{k \in K_p| k \in [0,\pi/2]\}$.
For technical convenience, we also set $\widetilde{K}^+_p \coloneq \{k \in K_p| k \in (0,\pi/2]\}$, which removes the term $k=0$ from $K^+_p$ if it exists, and likewise $\bar{K}_p^+ 
    \coloneq \{k \in K_p| k \in (0,\pi/2)\}$.
For $k \in \widetilde{K}^+_p$, we write the four-component Nambu spinor as
\begin{equation} \label{eq:def_nambu_spinor_4}
    \Psi_{\mathrm{B},k} = \left( c_k, c_{-k}^\dagger, c_{k+\pi}, c_{-k+\pi}^\dagger \right)^T.
\end{equation}
We also denote by $\Psi_{\mathrm{B},0}  = (c_0, c_\pi^\dagger)^T$ for $k=0$.
The operators $\log T_a$ are then written in the BdG form with $4 \times 4$ Hamiltonians,
\begin{equation} \label{eq:BdG_form_brickwall_single_layner}
\begin{aligned}
    \log T_a =& \frac{n}{2}x_u + \sum_{k \in \widetilde{K}_p^+} w_k \Psi_{\mathrm{B},k}^\dagger \mathcal{H}_a(k) \Psi_{\mathrm{B},k} \\
    &+\delta_{0}\left(\Psi_{\mathrm{B},0} ^\dagger {\mathcal{H}}_a(0)\Psi_{\mathrm{B},0} +x_u \right),
\end{aligned}
\end{equation}
where $ x_u \coloneq ({1}/{2})\log\!\left(1 - {e_u}/{e_{\mathrm{H}}}\right)$, and for $k \in \widetilde{K}^+_p$,
\begin{equation} \label{eq:BdG_Hamiltonian_brick-wall}
\begin{aligned}
    \mathcal{H}_a(k) \coloneqq x_u \Big[ \cos\alpha \, \mathbb{I} Z - \cos k \, Z Z 
           + \sin\alpha \sin k \, Z  Y \\
           - (-1)^a \big( \sin k \,Y  Z + \sin\alpha \cos k \, Y  Y\big) \Big],
\end{aligned}
\end{equation}
with $\alpha$ defined by ${2d}/{(d^2+1)} = \cos\alpha$, and for $k=0$,
\begin{equation}
\begin{aligned}
   {\mathcal{H}}_a(0) & \coloneqq   x_u 
    \begin{pmatrix}
        \cos\alpha - 1 & (-1)^a \sin\alpha \\
        (-1)^a \sin\alpha & -\cos\alpha - 1
    \end{pmatrix}.
\end{aligned}
\end{equation}
where the factor $w_k$ is defined by $w_k=1$ for all $0<k<\pi/2$ and $w_{\pi/2}=1/2$, and $\delta_0$ is defined by $\delta_0=1$ if $0 \in K_p^+$ and $\delta_0=0$ otherwise.
The derivation for the BdG form in Eq.~\eqref{eq:BdG_form_brickwall_single_layner} is written in Appendix~\ref{app:sec_BdG_brick_wall}.

Since the quadratic operators $\Psi_{\mathrm{B},k}^\dagger \mathcal{H}_a(k) \Psi_{\mathrm{B},k}$ commute for different wave numbers $k \neq k' \in K_p^+$, the exponential of the sum factorizes into a product of exponentials:
\begin{align} \label{eq:Ta_product_form}
    T_a = e^{x_u\left(\frac{n}{2}+\delta_{0}\right)} 
    \prod_{k \in K_p^+} \exp \left( w_k\Psi_{\mathrm{B},k}^\dagger \mathcal{H}_a(k) \Psi_{\mathrm{B},k} \right).
\end{align}
It is important for the later analysis that we have the relation between $\mathcal{H}_0(k)$ and $\mathcal{H}_1(k)$ as, for $k \in \widetilde{K}_p^+$,
\begin{equation} \label{eq:relation_H_1_H_0}
    \mathcal{H}_1(k) = Z_1 \mathcal{H}_0(k) Z_1,
\end{equation}
where we have defined $A_1 \coloneqq  A \otimes \mathbb{I}$ for an operator $A$.

\subsection{Exact diagonalization}
\label{subsec:ecaxt_diagonalization}
We now diagonalize the second-moment operator $M_{\mathrm{B},u}$. 
To this end, we diagonalize the BdG Hamiltonians $H_a(k)$.
Although the matrix of $k=0$ is a $2 \times 2$ matrix and can be diagonalized easily, each wave number $k \in \widetilde{K}_p^+$ gives a $4 \times 4$ matrix, which may not be straightforward to diagonalize. We can still obtain the exact diagonalization for these $4 \times 4$ matrices as follows.

The product of the two exponentials can be written as a single exponential by 
the following lemma.

\begin{lem}[Two exponentials as one] \label{lemma:product_exponentials_nambu_spinors}
    Let $k \in K_p^+$ such that $k\neq 0,\pi/2$ and $\Psi_{\mathrm{B},k}$ be the Nambu spinor defined in Eq.~\eqref{eq:def_nambu_spinor_4}.
    If $4 \times 4$ matrices $A$ and $B$  satisfy the relation $e^Ae^B=e^C$ for some $C$, then we have that 
    \begin{equation}
    e^{\Psi_{\mathrm{B},k}^\dagger A \Psi_{\mathrm{B},k}} e^{\Psi_{\mathrm{B},k}^\dagger B \Psi_{\mathrm{B},k}} = e^{\Psi_{\mathrm{B},k}^\dagger C \Psi_{\mathrm{B},k}}.
\end{equation}
\end{lem}
\noindent
Lemma~\ref{lemma:product_exponentials_nambu_spinors} is proven in Appendix~\ref{app:technical_lemma}.

From the above lemma, for each $k$, we obtain 
\begin{equation}
     M_{\mathrm{B},u}= e^{x_u \left({n}+2\delta_{0}\right)} \prod_{k \in K_p^+} \exp\left( w_k\Psi_{\mathrm{B},k}^\dagger C_k \Psi_{\mathrm{B},k} \right),
\end{equation}
where $4 \times 4$ matrices $C_k$ satisfy
\begin{equation} \label{eq:exp_c_k}
\begin{aligned}
    \exp\left(C_k\right) &= \exp\left({w_k Z_1 \mathcal{H}_0(k) Z_1}\right) \exp\left({w_k\mathcal{H}_0(k)}\right)\\
    &=\left[Z_1 \exp\left({w_k\mathcal{H}_0(k)}\right) \right]^2
\end{aligned}
\end{equation}

To diagonalize $M_{\mathrm{B},u}$, it is sufficient to diagonalize the non-Hermitian matrices $\exp(C_k)$ for $k \in K^+_p$. This is because if we have $\exp(C_k)=V_kD_kV^{-1}_k$ for some diagonal matrix $D_k$ and invertible matrix $V_k$, it yields $C_k=V_k \log(D_k)V_k^{-1}$, and 
\begin{equation}
\begin{aligned}
    M_{\mathrm{B},u}
    &=e^{x_u \left({n}+2\delta_{0}\right)} \prod_{k \in K_p^+}\exp\left( w_k\widetilde{\Phi}_{\mathrm{B},k} (\log D_k) {\Phi}_{\mathrm{B},k}\right),
    \end{aligned}
\end{equation}
where we have defined the right and left vectors of quasi-particles ${\Phi}_{\mathrm{B},k} \coloneq V_k^{-1} \Psi_{\mathrm{B},k}$ and $\widetilde{\Phi}_{\mathrm{B},k} \coloneq \Psi_{\mathrm{B},k}^\dagger V_k$, respectively.

We write the elements of the right quasiparticle vector $\Phi_{\mathrm{B},k}$ with annihilation-like operators $\eta$  as $\Phi_{\mathrm{B},k} = (\eta_{k,1}, \eta_{k,2}, \eta_{k,3}, \eta_{k,4})^\trans$.
Because the diagonalizing matrix $V_k$ is not unitary in general,  the left quasiparticle vector $\widetilde{\Phi}_{\mathrm{B},k}$ is not simply the Hermitian conjugate of $\Phi_{\mathrm{B},k}$, and we write the elements as $\widetilde{\Phi}_{\mathrm{B},k} = (\bar{\eta}_{k,1}, \bar{\eta}_{k,2}, \bar{\eta}_{k,3}, \bar{\eta}_{k,4})$. It is straightforward to show that, for an invertible matrix $V_k$, they satisfy the canonical anticommutation relations, in the sense that for all $k \in \bar{K}_p^+$ and $i,j\in\{1,2,3,4\}$,
\begin{equation} \label{eq:generalized_anticommutation_quasiparticle_vectors}
    \{\eta_{k,i}, \bar{\eta}_{k,j}\} = \delta_{i,j}, \qquad \{\eta_{k,i}, \eta_{k,j}\} = \{\bar{\eta}_{k,i}, \bar{\eta}_{k,j}\} = 0.
\end{equation}
With this relation, we have
\begin{equation} \label{eq:diagonal_expanded}
\begin{aligned}
    w_k\widetilde{\Phi}_{\mathrm{B},k} (\log D_k) {\Phi}_{\mathrm{B},k} 
    =  \sum_{j=1}^4 w_k(\log \lambda_{k,j}) \bar{\eta}_{k,j} \eta_{k,j},
\end{aligned}
\end{equation}
where $\lambda_{k,i}$ are the eigenvalues of $D_k$.

We define the non-Hermitian fermionic number operators as $\hat{n}_{k,j} \coloneq \bar{\eta}_{k,j} \eta_{k,j}$. 
From Eq.~\eqref{eq:generalized_anticommutation_quasiparticle_vectors}, they satisfy $[\hat{n}_{k,i}, \hat{n}_{k,j}] = 0$ and $\hat{n}_{k,j}^2 = \hat{n}_{k,j}$.  
We thus have, for each $k \in \bar{K}_p^+$,
\begin{equation}
\begin{aligned}
    \exp\left( \sum_{j=1}^4 (\log \lambda_{k,j}) \hat{n}_{k,j} \right) = \prod_{j=1}^4 \left( \lambda_{k,j} \right)^{\hat{n}_{k,j}}.
\end{aligned}
\end{equation}
Likewise, for $k=\pi/2$, we have
\begin{equation} \label{eq:eigenvalues_lambda_k_boundary}
\begin{aligned}
    \exp\left( \sum_{j=1}^4 \frac{1}{2}(\log \lambda_{k,j}) \hat{n}_{k,j} \right) 
    &= \prod_{j=1}^2 \left( \lambda_{k,j} \right)^{\hat{n}_{k,j}},
\end{aligned}
\end{equation}
where we have used the fact that only two wave numbers $k=\pi/2$ and $k=-\pi/2$ couple, and $\hat{n}_{k,1}$ and $\hat{n}_{k,2}$ are redefined by the corresponding occupation numbers. (see Appendix~\ref{app:subsec_k=pi/2} for detail).

We define a right vacuum state vector $|R_0\rangle$ and a left vacuum state $\langle L_0|$ such that they are annihilated as follows: for all $k\in K^+_p$ and $j\in\{1,2,3,4\}$,
\begin{equation}
    \eta_{k,j} \ket{R_0} = 0, \quad \bra{L_0} \bar{\eta}_{k,j} = 0.
\end{equation}
The right eigenvectors are obtained by applying some of the creation-like operators $\bar{\eta}_{k,j}$ to $\ket{R_0}$, and the corresponding eigenvalues of $M_{\textrm{B},u}$ are the products of the eigenvalues ${\lambda}_{k,j}$ of quasi-particle $\bar{\eta}_{k,j}$ (see Appendix~\ref{subsection:eigenvectors} for detail). The left eigenvectors are obtained similarly.
Their explicit eigenvalues are given later in Eq.~\eqref{eq:eigenvalues:moment_op_M_B_u}.

\subsection{Explicit formulas for the eigenvalues}
\label{sec:formula_evalues_brick_wall}

It remains to diagonalize the $4 \times 4$ matrix $\exp(C_k)=\left[Z_1 \exp\left(w_k \mathcal{H}_0(k)\right) \right]^2$, which yields the eigenvalues and eigenvectors of $M_{\mathrm{B},u}$.
To this end, we calculate
\begin{equation}
\begin{aligned}
    (\mathcal{H}_0(k))^2 
    = 2x_u^2 ( I - \bm{n}(k) \cdot \bm{\sigma}),
\end{aligned}
\end{equation}
where we have defined 
\begin{equation} \label{eq:def_bm_n_sigma}
    \begin{aligned}
        \bm{n}(k) &\coloneq (\cos\alpha \cos k, \cos\alpha \sin k, \sin\alpha),\\
        \bm{\sigma} &\coloneq(Z_1,Y_1,XX).
    \end{aligned}
\end{equation}
 Since $|\bm{n}|^2 =1$ and the pauli matrices in $\bm{\sigma}$ are mutually anti-commuting, the operator 
\begin{equation}
    P(k) \coloneqq \frac{1}{2}( I - \bm{n}(k) \cdot \bm{\sigma})
\end{equation}
is a projector.
By definition, we have that $(\mathcal{H}_0(k))^2=4 x_u^2P(k)$, and thus $[\mathcal{H}_0(k),P(k)]=0$. 
Since $\mathcal{H}_0(k)$ is Hermitian, its kernel is equal to that of $(\mathcal{H}_0(k))^2$, and $P$ is the projector onto the image of $\mathcal{H}_0(k)$.
Therefore, we have 
\begin{equation}
    (\mathcal{H}_0(k))^3 =4 x_u^2P(k)\mathcal{H}_0(k)=4 x_u^2\mathcal{H}_0(k).
\end{equation}
From the above relation, the higher power of $\mathcal{H}_0$  becomes simply
\begin{equation} \label{eq:H0_powers}
    (\mathcal{H}_0)^l = \begin{cases}
        (2x_u)^{l-1} \mathcal{H}_0 & (\text{odd } l \ge 1), \\[4pt]
        (2x_u)^{l-2} \mathcal{H}_0^2 & (\text{even } l \ge 2).
    \end{cases}
\end{equation}
The exponential of an operator is defined by its Taylor series, and we obtain
\begin{equation} \label{eq:exp_intermediate}
\begin{aligned}
    e^{w_k\mathcal{H}_0(k)} 
    &=\sum_{l=0}^\infty \frac{1}{l!}\left(w_k\mathcal{H}_0(k)\right)^l\\
    &= \frac{\cosh(\tilde{x}_{k})+1}{2}I + \frac{\sinh(\tilde{x}_{k})}{2{x}_{u}}\mathcal{H}_0(k) \\
    & \quad - \frac{\cosh(\tilde{x}_{k})-1}{2}\bm{n}(k) \cdot \bm{\sigma},
\end{aligned}
\end{equation}
where we have defined $\tilde{x}_{k} \coloneq 2w_k x_u$. 

We derive the eigenvalues of $\widetilde{M}_k \coloneqq Z_1 e^{w_k\mathcal{H}_0(k)}$, which immediately yields the desired ones of $\exp(C_k)$ as shown in Eq.~\eqref{eq:exp_c_k}. 
To this end, we employ the following lemma, which is proven in Appendix~\ref{app:technical_lemma}.
\begin{lem}[Eigenvalues as reciprocal pairs] \label{lem:orthogonal_reciprocal_pairs}
The eigenvalues of $4 \times 4$ matrix $\widetilde{M}_k = Z_1 e^{w_k\mathcal{H}_0(k)}$, for $k\in \bar{K}^+_p$, form reciprocal pairs: if $\lambda \neq 0$ is an eigenvalue, $\lambda^{-1}$ is also an eigenvalue.
\end{lem}

Let us consider $k \in \widetilde{K}_p^+$. From Lemma~\ref{lem:orthogonal_reciprocal_pairs}, the matrices $\widetilde{M}_k$ have four eigenvalues in reciprocal pairs $(\mu_0, \mu_0^{-1}, \mu_1, \mu_1^{-1})$. 
We define $\bar{\mu}_0 \coloneq  \mu_0 + \mu_0^{-1}$ and $\bar{\mu}_1 \coloneqq  \mu_1 + \mu_1^{-1}$. The traces of the matrix are related by 
\begin{equation} \label{eq:trace_M_k}
    \begin{aligned}
        \mathrm{Tr}(\widetilde{M}_k) &= \bar{\mu}_0 + \bar{\mu}_1, \\
        \mathrm{Tr}(\widetilde{M}_k^2) &= (\bar{\mu}_0)^2 + (\bar{\mu}_1)^2 - 4.
    \end{aligned}
\end{equation}
On the other hand, from Eqs.~\eqref{eq:BdG_Hamiltonian_brick-wall}, \eqref{eq:def_bm_n_sigma}, and \eqref{eq:exp_intermediate}, we obtain
\begin{equation} \label{eq:trace_I1}
\begin{aligned}
    \mathrm{Tr}(\widetilde{M}_k) 
    &= -2(\cosh(\tilde{x}_{k})-1)\cos\alpha \cos k,\\
    \mathrm{Tr}(\widetilde{M}_k^2) 
    &= 4\cosh(\tilde{x}_{k})\big(1 + (\cosh(\tilde{x}_{k})-1)\cos^2\alpha \cos^2 k \big).
\end{aligned}
\end{equation}

We can solve these equations for  $\mu_{i}$, which yield the eigenvalues of $M_{\mathrm{B},u}$.
These solutions yield the eigenvalues of $M_{\mathrm{B},u}$ that are given by, for each $p \in \{0,1\}$, 
\begin{equation} \label{eq:eigenvalues.M_Bu_main}
    \prod_{k \in K_p} (\lambda_k)^{n_k}
\end{equation}
where $\{n_{k}\}_{k\in K_p}$, for $n_{k} \in \{0,1\}$,  satisfies that $\sum_{k \in K_p}n_{k}$ is even, and 
\begin{equation} \label{eq:eigenvalues_lambda_main}
    \lambda_k=\left(
a_k\widetilde e_u
+
\sqrt{
a_k^2\widetilde e_u^2+1-\widetilde e_u
}
\right)^2,
\end{equation}
where $ \widetilde e_u={e_u}/{e_{\mathrm{H}}}$, $ a_k = ({d}/({d^2+1}))\cos{k}$, and we describe the derivations of Eqs.~\eqref{eq:eigenvalues.M_Bu_main} and \eqref{eq:eigenvalues_lambda_main} in Appendix~\ref{app:eigenvalues}.

From the above result, we find that while the eigenvalues of $M_{\mathrm{L},u}$ are always real, those of $M_{\mathrm{B},u}$ can be non-real due to their non-Hermiticity.
We note that 
\begin{equation}
\lambda_{\pi-k}=\left(a_k\widetilde{e}_u-\sqrt{a_k^2\widetilde{e}_u^2+1-\widetilde{e}_u}\right)^2=\lambda_k^* 
\end{equation}
when $a_k^2\widetilde{e}_u^2+1-\widetilde{e}_u<0$, where the eigenvalues come in complex conjugate pairs because $M_{a,u}$ for $a=L,B$ are real matrices.

\subsection{Exact spectral gap}
\label{sec:spectral_gap_brick}
We finally derive the spectral gap of the second-moment operator $M_{\mathrm{B},u}$, for $u$-strucrtured brick-wall random circuits and prove that $u$-structured brick-wall circuits can be more random with respect to spectral gap than brick-wall Haar random circuits.

The absolute value of the eigenvalues $| \lambda_{k}|$ satisfy $| \lambda_{k}| \leq 1$ for all $k\in K_p$, and in particular $| \lambda_{k}| < 1$ for an entangling gate $u$. Then, the largest eigenvalue is obtained by choosing $n_k=0$ for all $k \in K_p$ in Eq.~\eqref{eq:eigenvalues.M_Bu_main}. It implies that the largest eigenvalue is $1$, as expected for a moment operator.

We next derive the second-largest eigenvalue.
From the expression for $\lambda_{k}$, we find that $|\lambda_{k}|$ is monotonically decreasing in $k$ for $0 \leq k \leq {\pi}$. 
Therefore, the second-largest eigenvalue in absolute value in the even and odd parity sectors are $| \lambda_{\frac{\pi}{n}}|^2$ and $|\lambda_{0}\lambda_{\frac{2\pi}{n}}|$, respectively. 
We note that $\lambda_{0}$ is a real number for any integer $d \geq 2$ and real number $0 \leq e_u \leq {(d^2-1)}/{d^2}$. 
We then obtain the spectral gap as 
\begin{equation} \label{eq:second_largest_BW}
1-\max_{i,j \in K_p}\left( |\lambda_i \lambda_j| \right) = 1-\max\left( | \lambda_{\frac{\pi}{n}}|^2, |\lambda_{0}\lambda_{\frac{2\pi}{n}}| \right).
\end{equation}
With Eqs.~\eqref{eq-def-lambda-k} and \eqref{eq:lambda_k=0_pi/2}, Eq.~\eqref{eq:second_largest_BW} is the exact formula for the spectral gap.

We further compute the right-hand side in Eq.~\eqref{eq:second_largest_BW}. 
When $\lambda_{\frac{2\pi}{n}}$ is a real number, $\lambda_{\frac{\pi}{n}}$ is also real. 
In this case,  we have that $(\lambda_{ \frac{\pi}{n}})^2 \geq \lambda_{0}\lambda_{\frac{2\pi}{n}}$ due to the concavity of $\log\lambda_k$, and
 the gap $1-(\lambda_{\frac{\pi}{n}})^2$ monotonically increases with $e_u$.
We remark that $\lambda_{\frac{2\pi}{n}}$ is real at least in the regime $0\leq e_u\leq e_{\mathrm{H}}$.
On the other hand, $\lambda_{k}$  can be a complex number, for some $0 < k \leq {\pi}/{2}$, if $e_u > e_{\mathrm{H}}$. 
When $\lambda_{k}$  is not real,  we have $|\lambda_{k}|=\frac{e_u}{e_{\mathrm{H}}}-1$, which does not depend on $k$. 
Additionally, when $\lambda_{\frac{\pi}{n}}$ is not a real number, so $\lambda_{\frac{2\pi}{n}}$ is, 
and in this case we have that 
$\max\left( | \lambda_{\frac{\pi}{n}}|^2, |\lambda_{0}\lambda_{\frac{2\pi}{n}}| \right)=\lambda_{0}|\lambda_{\frac{2\pi}{n}}|=\lambda_{0} \left( {e_u}/{e_{\mathrm{H}}}-1 \right)$.
Here, we have used that $\lambda_{0}$ is a real number for any integer $d \geq 2$ and $0 \leq e_u \leq ({d^2-1})/{d^2}$, which is true under the solvability condition and Eq.~\eqref{eq:eu-gu-upper-boundary}.
Therefore, if $\lambda_{\frac{\pi}{n}}$ is not real, the spectral gap of $M_{\mathrm{B},u}$ is
\begin{equation}
    1-\lambda_{0} \left( \frac{e_u}{e_{\mathrm{H}}}-1 \right),
\end{equation}
which, in contrast, decreases monotonically with $e_u$.

We finally prove more randomness with respect to the spectral gap for structured brick-wall random circuits, as follows.
When $\lambda_{\frac{ \pi}{n}}$ is real, the spectral gap also increases monotonically at least for $e_u \leq e_{\mathrm{H}}$.
As a special case corresponding to Haar-random gates, the spectral gap is obtained by setting $e_{\mathrm{H}}=e_u$ in Eq.~\eqref{eq:second_largest_BW}, yielding
\begin{equation}
    1-\left(
    \frac{d}{d^2+1}
    \right)^4\left(2\cos\frac{2\pi}{n}+2\right)^2,
\end{equation}
which agrees with the spectral gap derived in Ref.~\cite{deneris2024exactspectralgapsrandom}.
When $\lambda_{\frac{ \pi}{n}}$ is a non-real number, we have that
\begin{equation}
1-\lambda_{0}\left(\frac{e_u}{e_{\mathrm{H}}}-1 \right) \geq 1-\left(
    \frac{d}{d^2+1}
    \right)^4\left(2\cos\frac{2\pi}{n}+2\right)^2,
\end{equation}
where we have used $e_{\mathrm{H}} < e_u \leq  ({d^2-1})/{d^2}$ and $n\geq 4$. Therefore, the gap of $u$-structured brick-wall circuits is larger than that of brick-wall Haar random circuits if and only if $e_u>e_{\mathrm{H}}$.

\begin{figure*}[t]
  \centering
  \includegraphics[width=18cm]{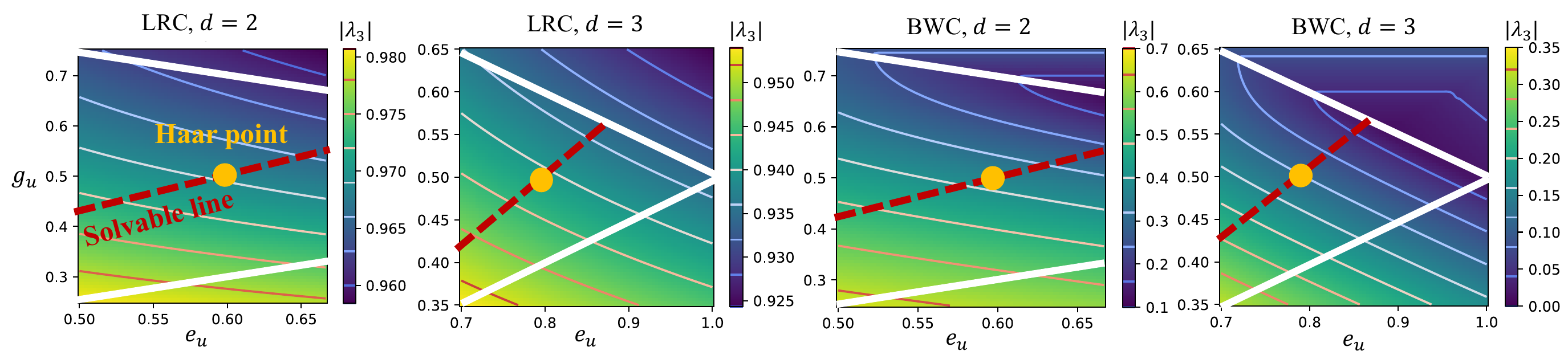}
\caption{Numerical results on the second-largest eigenvalues in absolute value $|\lambda_3|$  of $M_{L,u}$ (local) and $M_{B,u}$ (brick-wall) for general $e_u$ and $g_u$ in qubit system ($d=2$) and qutrit system ($d=3$) in the case of $n=14$. 
  The x-axis and the y-axis 
  refer 
  to the entangling power $e_u$ and the gate typicality $g_u$, respectively.
  We derive analytical solutions of them on the solvable lines (the broken red lines) satisfying the solvability condition. At the 
  Haar points (the orange dots), the second-moment operators of structured random circuits are equal to 
  those of the Haar random circuits with the same architecture, and the 
  points are on the solvable lines.
  Only the values between the thick white lines are feasible for $e_u$ and $g_u$, as discussed in Section~\ref{entangling_power_gate_typicality} and also in Ref.~\cite{PhysRevResearch.2.043126}.
  } 
  \label{fig:numerics_BW_second-largest}
\end{figure*}

\section{Numerics}
Along with the analytical solutions, we numerically calculate the second-largest eigenvalues in absolute value $|\lambda_3|=1-\Delta_{\nu}$ of $M_{L,u}$ and $M_{B,u}$, as shown in Fig.~\ref{fig:numerics_BW_second-largest}. 
The solvability condition is illustrated as the solvable 
lines there, and we confirm that the analytical results agree with the numerical results on the lines. Figure~\ref{fig:numerics_BW_second-largest} clearly illustrates our main statement that the spectral gap becomes larger than that of Haar points when $e_u>e_{\mathrm{H}}$, where $e_{\mathrm{H}}={3}/{5}$ for $d=2$ and $e_{\mathrm{H}}={4}/{5}$ for $d=3$.

We find an interesting difference between local random 
circuits and brick-wall circuits as follows. In Figs.~\ref{fig:numerics_BW_second-largest} (a) and (b) in the case of local random circuits, the spectral gap monotonically increases in $g_u$, as proven in Theorem~\ref{thm:more_randomness_loca_random}.
However, in Figs.~\ref{fig:numerics_BW_second-largest} (c) and (d) in the case of brick-wall random circuits,
it reaches the minimal value, which is close to the line $e_u=2-2g_u$ (the thick upper white line shown in each plot), when we change $g_u$ with fixed $e_u$. The two-qudit gates on the line $e_u = 2-2g_u$ are the dual-unitary gates, which serve as a toy model of chaotic dynamics \cite{bertini2025exactlysolvablemanybodydynamics}. We leave it as an open problem to determine whether a dual-unitary brick-wall random circuit achieves the largest spectral gap in the structured brick-wall random circuits with a given entangling power.

\section{Applications} 
We suggest that our work may open up a number of compelling applications.
{We can apply the enhancement of randomness  in terms of the second moment to the followings:  }
{{(1) \emph{Reduction of circuit depth for randomized benchmarking}.}}
\emph{Randomized benchmarking} (RB) \cite{PhysRevLett.106.180504} is a widely used technique for estimating the quality of quantum gates in quantum computers. {Moreover, random unitary circuits consisting of random local gates serve as tools for assessing the error rate, known as the filtered RB, and the circuit depth for realizing it is upper-bounded by the inverse of the spectral gap of the second-moment operator in a sufficiently large depth \cite{heinrich2022randomized}. }
Combining it with our enhanced spectral 
gaps, we can improve the required circuit depth for the {filtered RB while maintaining a theoretical performance guarantee}. 
{{(2) \emph{Improved bound on unitary 2-designs in shallow random circuits}.}}
Unitary $2$-designs have a range of applications, such as computational 
advantage \cite{hangleiter2018anticoncentration, PRXQuantum.3.010333}, quantum channel fidelity estimation \cite{PhysRevA.80.012304}, random coding \cite{roy2009unitary, PhysRevX.11.031066}, and randomized measurement \cite{bae2019linking, elben2023randomized}. Recently, 
Haar random 
circuits have been shown 
to form approximate unitary designs in depth $\log{n}$ by the technique of gluing small random circuits \cite{schuster2024random, laracuente2024approximate}. Since the structured random circuits can have better upper bounds on the depth generating approximate unitary $2$-designs, combining it with the technique by gluing small structured random circuits, we can improve the approximate $2$-design time.
{{(3) \emph{Exact result on quantum chaotic dynamics}.}}
More physically motivated applications of our results 
manifest themselves by the connection between the second-moment operator and indicators of quantum chaos, such as 
\emph{out-of-time ordered correlators} 
(OTOC) and the R\'enyi-2  operator entanglement entropy \cite{roberts2017chaos, hosur2016chaos, PhysRevB.95.094206, fisher2023random}. Because we derive the eigenvalues and eigenvectors of the second-moment operators satisfying the solvability condition, in principle,  one can
obtain closed formulas for these indicators in structured random 
circuits satisfying the solvability condition.

\section{Discussion and conclusion.} 
In our work, we have developed a theory of structured random circuits by the technique of exact diagonalization and have proven that random circuits with non-Haar random gates can be more random than those with Haar random gates. {We expect that the solvable random gates introduced in this work will be of independent interest for the study of quantum many-body dynamics, where exact results on real-time evolution play a central role, as exemplified by the case of dual-unitary gates \cite{bertini2025exactlysolvablemanybodydynamics}.}
Moreover, our work also applies beyond the solvable free-fermionic regime, since Theorem~\ref{thm:more_randomness_loca_random} does not rely on the solvability condition.
Also,  more randomness with respect to positivity in structured local random circuits is true for an arbitrary spatial dimension, not only a one-dimensional system.

The additional structure improves the convergence rate to a unitary $2$-design ensemble, which will be particularly important when we use large-scale quantum computers. Since we have considered the second-moment, the Haar random single-qudit gates in the structured random gates can be replaced by any gates forming unitary 2-design locally, keeping our results the same.

While we have focused on the diagonalization of moment operators in the main text, we also derive the bounds on the frame potential in structured random circuits by the statistical mechanical mapping \cite{PhysRevX.8.021014, hunter2019unitary}, as shown in Appendix~\ref{sec-frame-potential}.

We note that Ref.~\cite{deneris2024exactspectralgapsrandom} has also derived the spectral gap of brick-wall Haar random circuits, and our result at the Haar point of $e_u=e_{\mathrm{H}}$ and $g_u=g_{\mathrm{H}}$ agrees with it.
Also, the spectra of matchgate brick-wall  circuits have been considered in Ref.~\cite{Richelli_2024}, and we expect that our results on the diagonalization of brick-wall circuits are also applicable to matchgate 
unitary circuits.
Within the context of entanglement dynamics, phenomena displaying a faster decay of purity in certain structured random circuits, which is similar to more randomness in our setting, have been observed in the early works 
 on random circuits Refs.~\cite{vznidarivc2007optimal, PhysRevA.78.032324}
(see Appendix~\ref{sec-Related_works} for further details on related works). 

We also note that, subsequent to our preprint of this work, a few works appeared that study a similar line of research from different perspectives \cite{kong2024convergenceefficiencyquantumgates, belkin2025absencecensoringinequalitiesrandom, riddell2025quantumstatedesignsminimally}, which connect certain fixed gates, such as dual-unitary gates, with the fast generation of randomness measured by unitary or state designs.

Our results open up numerous opportunities for future work. It would be interesting to show more randomness of structured random circuits in terms of higher-moment operators, as well as other architectures, for example,  higher dimensional brick-wall circuits. Other future directions are finer analyses of the applications we have discussed above.
Also, investigating the randomness in Brownian Hamiltonian dynamics \cite{RandomHamiltonians, jian2023linear, PhysRevLett.123.080601} with an additional structure similar to this work would be a physically interesting direction. At the end of such a program stands an idea of how to best create randomness with {the} smallest possible circuit complexity. We hope that 
this work helps to identify 
the fundamental limits of 
what notions of randomness one can achieve with reasonable constraints on circuit complexity.

\section*{Acknowledgements.}
We thank Ingo Roth, Janek Denzler, Tomohiro Yamazaki, Christian Bertoni, and Marko Žnidarič for useful discussions.
The Berlin team acknowledges the support by the German  Federal Ministry of Research, Technology and Space 
(PhoQuant, DAQC, MuniQC-Atoms), the Munich Quantum Valley, the DFG (CRC 183, FOR 2724, SPP 2514), ML4Q, Berlin Quantum and the European Research Council (DebuQC). H.~K. is supported by JSPS KAKENHI Grants No.\ JP23K25783, No.\ JP23K25790, and MEXT KAKENHI Grant-in-Aid for Transformative Research Areas A “Extreme Universe” (KAKENHI Grant No.\ JP21H05191).
Y.~M. is supported by JSPS KAKENHI Grant No.\ JP23KJ0421.
N.~Y. wishes to thank JST PRESTO No.\ JPMJPR2119, JST ASPIRE Grant Number JPMJAP2316,
and the support from IBM Quantum.
This research is funded in part by the
Gordon and Betty Moore Foundation’s EPiQS Initiative,
Grant GBMF8683 to T.S.
This work was supported by JST Grant Number JPMJPF2221, JST ERATO Grant Number JPMJER2302, and JST CREST Grant Number JPMJCR23I4, Japan. 

\section*{Data availability}
The code used to generate Fig.~2 is publicly available in a GitHub repository~\cite{DataCode2026}.

\appendix

\section{Related works and comparison}
\label{sec-Related_works}

References ~\cite{deneris2024exactspectralgapsrandom, rampp2024hayden} also derive the exact spectral gap of one-dimensional brick-wall Haar random circuits, and the former reference considers Haar measures on various compact groups. In this work, our interests are in non-Haar random gates, which exhibit more randomness, and we derive all the eigenvalues and eigenvectors.
References~\cite{PhysRevX.7.021006, haferkamp2023momentsrandomquantumcircuits, metger2024simpleconstructionslineardepthtdesigns, chen2024incompressibility} also consider other kinds of structured random walks to obtain better upper bounds on convergence time than the existing upper bounds on those of Haar random circuits. 
We note that, for our purpose, the upper bounds alone are not sufficient to imply more randomness because the upper bounds alone do not determine which moment operator has the larger spectral gap compared to the other.

The early works on random circuits by \v{Z}nidari\v{c}~\cite{vznidarivc2007optimal, PhysRevA.78.032324} have considered a type of structured random circuits in qubit systems, in the context of entanglement dynamics. 
Reference~\cite{vznidarivc2007optimal} numerically observed that the purity of a quantum state generated by a random circuit with a specific gate structure can converge to the purity of Haar random states faster than Haar random circuits (see also Ref.~\cite{PhysRevX.11.031019}).
Reference~\cite{PhysRevA.78.032324} analytically diagonalizes the second moment of one-dimensional Haar random local circuits and obtains the asymptotic formula for the spectral gap of local random circuits with all-to-all connectivity and specific gate structure. 
{Also, Gao \textit{et al.}~\cite{Gao_2024} and Ware \textit{et al.}~\cite{ware2023sharp} have considered a structured random gate with a fixed interacting gate and single-qubit Haar random gates in the context of the quantum computational advantage, and the latter numerically computed the spectral gaps of the second-moment operators, showing that Haar-random gates do not extremize the spectral gap. }

In this work, we focus on randomness generation within the framework of unitary designs and extend the diagonalization method to a broad family of random gates satisfying the solvable condition, as well as the one-dimensional brick-wall architecture, not only the local circuit architecture.
As far as we know, our work is the first to derive exact spectral gaps of the second moment of one-dimensional non-Haar random circuits, which works for an arbitrary number of qudits and can exceed the spectral gap of Haar random circuits with the corresponding circuit architecture. Also, Ref.~\cite{bensa2022two} has conjectured on the exact spectral gap of structured dual-unitary random circuits, and our exact diagonalization for the structured dual-unitary random circuits satisfying the solvable condition can support the conjecture. Moreover, our work further examines regimes outside the solvable free-fermionic framework, namely frustration-free interacting fermionic chains, and extends the analysis beyond one-dimensional architectures, as mentioned in Section~\ref{sec-more-randomness-positivity-LRC}.

\section{Solvability condition}
\label{app:solvability_condition_qubit}

We derive Eq.~\eqref{eq:Wu_Pauli_final} as follows. We expand the operator $W^u$ as 
\begin{equation}
\label{eq:Wu_matrix_from_Pauli}
\begin{aligned}
W^u
&=
a_0\,\mathbb{I}\mathbb{I}
+a_1\bigl(Z\mathbb{I}+\mathbb{I}Z\bigr)
+a_2\,XX
+a_3\,YY
+a_4\,ZZ
\\
&=
\begin{pmatrix}
a_0+2a_1+a_4 & 0 & 0 & a_2-a_3 \\[6pt]
0 & a_0-a_4 & a_2+a_3 & 0 \\[6pt]
0 & a_2+a_3 & a_0-a_4 & 0 \\[6pt]
a_2-a_3 & 0 & 0 & a_0-2a_1+a_4
\end{pmatrix}.
\end{aligned}
\end{equation}
for qubit-Pauli operators $\{\mathbb{I},X,Y,Z\}$ and some real coefficients $a_i$ for $i \in \{0,1,2,3,4\}$.
By comparing the matrix elements with the
representation in Eq.~\eqref{eq:Wu_matrix_representation},
we obtain that
\begin{equation}
\begin{aligned}
a_0 &= 1 - \frac{e_u}{4}\frac{d^2+1}{d^2-1}- \frac{g_u}{2}, \\
a_1 &= -\frac{d e_u}{2(d^2-1)}, \\
a_2 &= \frac{e_u+2g_u}{4}, \\
a_3 &=  -\frac{e_u-2g_u}{4},\\
a_4 &=\frac{g_u}{2} - \frac{e_u}{4}\frac{d^2+1}{d^2-1},
\end{aligned}
\end{equation}
which yields Eq.~\eqref{eq:Wu_Pauli_final} with the definitions of $e_{\mathrm{H}}$ and $g_{\mathrm{H}}$.

We next derive the solvability condition for a two-qubit unitary gate in Eq.~\eqref{eq:solvability_condition_qubit}.
For a two-qubit unitary gate $u$, its dual operator $\Tilde{u}$ is defined by reshuffling the indices as 
\begin{align}
    \bra{i}\otimes \bra{j}\Tilde{u}\ket{k}\otimes \ket{l}=\bra{i}\otimes \bra{k}u \ket{j}\otimes \ket{l}.
\end{align}
With the parametrization $u=e^{-\frac{i}{2}(\theta_x XX+\theta_y YY+\theta_z ZZ)}$ \cite{PhysRevA.63.062309}
we can express $\bra{I,S} u^{\otimes 2,2} \ket{IS}$ in terms of its dual operator as $\bra{I,S} u^{\otimes 2,2} \ket{I,S}= \Tr (\Tilde{u}\Tilde{u}^{\dagger}\Tilde{u}\Tilde{u}^{\dagger})$, where we have defined 
\begin{equation}
\label{eq:two_qubit_gate}
\begin{gathered}
\mathsf c_{\pm}\coloneqq \cos(\theta_x/2\pm\theta_y/2),
\qquad
\mathsf s_{\pm}\coloneqq \sin(\theta_x/2\pm\theta_y/2),
\\[2mm]
\widetilde{u}=
\begin{pmatrix}
e^{-i\theta_z/2}\mathsf c_{-} & 0 & 0 & e^{i\theta_z/2}\mathsf c_{+} \\
0 & -i e^{-i\theta_z/2}\mathsf s_{-} & -i e^{i\theta_z/2}\mathsf s_{+} & 0 \\
0 & -i e^{i\theta_z/2}\mathsf s_{+} & -i e^{-i\theta_z/2}\mathsf s_{-} & 0 \\
e^{i\theta_z/2}\mathsf c_{+} & 0 & 0 & e^{-i\theta_z/2}\mathsf c_{-}
\end{pmatrix}.
\end{gathered}
\end{equation}
By straightforward calculation, we obtain that
\begin{equation} \label{eq:matrix_elements_qubit_ISIS}
\begin{aligned}
    \bra{I,S} u^{\otimes 2,2} \ket{I,S}&=4(1+\cos^2(\theta_x)\cos^2(\theta_y)\\
    &+\cos^2(\theta_y)\cos^2(\theta_z)+\cos^2(\theta_z)\cos^2(\theta_x)),\\
\end{aligned}
\end{equation}
and
\begin{equation} \label{eq:matrix_elements_qubit_SISI}
\begin{aligned}
    \bra{I,S} u^{\otimes 2,2} \ket{S,I}
    &=\bra{I,S} (u\cdot \textrm{SWAP})^{\otimes 2,2} \ket{I,S} \\
&=4(1+\sin^2(\theta_x)\sin^2(\theta_y)\\
&+\sin^2(\theta_y)\sin^2(\theta_z)+\sin^2(\theta_z)\sin^2(\theta_x)), 
\end{aligned}
\end{equation}
where we have used $\textrm{SWAP}=e^{-\frac{\pi}{4}i(XX+YY+ZZ)}$ up to a phase factor.
From Eqs.~\eqref{eq:matrix_elements_qubit_ISIS} and \eqref{eq:matrix_elements_qubit_SISI} with Eq.~\eqref{eq:matrix_element_u_ISIS}, the entanglement properties $e_u$ and $g_u$ are written as
\begin{equation} \label{eq:e_u_g_u_theta}
\begin{aligned}
e_u
=& \frac{2}{3}\Big[
\sin^2\theta_x \cos^2\theta_y
+\sin^2\theta_y \cos^2\theta_z
+\sin^2\theta_z \cos^2\theta_x
\Big],\\
g_u
=&
\frac{1}{3}\Big[
\sin^2(\theta_x)
+\sin^2(\theta_y)
+\sin^2(\theta_z)
\Big].
\end{aligned}
\end{equation}
We note that Eq.~\eqref{eq:e_u_g_u_theta} has already been derived in Ref.~\cite{PhysRevResearch.2.043126}.
The solvability condition~\eqref{eq-SM-solvable-condition} can then be rephrased as 
\begin{equation}
    f(\theta_x, \theta_y)+f(\theta_y, \theta_z)+f(\theta_z, \theta_x)=0,
\end{equation}
for the function
\begin{equation}
    f(x_1,x_2) =\sin^2(x_1)\left[\cos^2(x_2)-\frac{3}{5}\right].
\end{equation}

\section{Fermionic representation of brick-wall circuits}

\subsection{Free fermion mapping}

The second-moment operator of $u$-structured brick-wall random circuits is given by $M_{\mathrm{B},u}=T_1T_0$.
Under the solvability condition, we map $\log T_a$ to a fermionic chain via the Jordan-Wigner transformation, and we obtain
\begin{equation}
\label{eq:log_T_fermion}
\begin{aligned}
\log T_a
&= x_u \Bigg[
A + B N
- \sum_{j \equiv a \!\!\!\pmod 2} h_{j,j+1}
\Bigg]\\
&= x_u \Bigg[
A + B N
- \sum_{j=1}^{n}
\frac{1+e^{i\pi(j-a)}}{2}\,
h_{j,j+1}
\Bigg],
\end{aligned}
\end{equation}
where $ x_u = ({1}/{2})\log\!\left(1 - {e_u}/{e_{\mathrm{H}}}\right)$, $A \coloneqq {n(d-1)^2}/({2(d^2+1)})$, $B \coloneqq {2d}/{(d^2+1)}$, 
and
\begin{equation}
\label{eq:local_fermion_term}
h_{j,j+1}
\coloneqq
c_j^\dagger c_{j+1}
+ c_{j+1}^\dagger c_j
+ e_{\mathrm{H}}
\left(
c_j^\dagger c_{j+1}^\dagger
+ c_{j+1} c_j
\right).
\end{equation}
Here, $N=\sum_{i=1}^nc_i^\dagger c_{i}$ is the number operator and the summation over $i$ respects the parity-dependent boundary condition $c_{n+1} = -(-1)^N c_1$. 
Concretely, the boundary term connecting site $n$ and site $1$ is determined by the parity sector to which the state belongs, as in Section~\ref{section:exact_XY_model}.
For $0<e_u<e_{\mathrm{H}}$, $x_u$ a negative number. For $e_u>e_{\mathrm{H}}$, it becomes a complex number, and we adopt the branch 
\begin{equation}x_u =\frac{1}{2}\log\!\left(\frac{e_u}{e_{\mathrm{H}}}-1\right)+i\pi/2.
\end{equation}

Since the operators $\log T_a$ are $2$-site translation invariant, we transform the fermionic operators into momentum space using the discrete Fourier transform, and the most right-hand side of  Eq.~\eqref{eq:log_T_fermion} becomes
\begin{equation}
\label{eq:log_T_k_space}
\begin{aligned}
&x_u \bigg[
A
+ \sum_{k\in K_p}
\bigg\{
(B-\cos k)c_k^\dagger c_k  
+
\bigg[
-\frac{i e_{\mathrm H}}{2}\sin k\,
c_k^\dagger c_{-k}^\dagger
\\
&
+(-1)^a
\left(
\frac{i}{2}\sin k\,
c_k^\dagger c_{k+\pi}
+
\frac{e_{\mathrm H}}{2}\cos k\,
c_k^\dagger c_{-k+\pi}^\dagger
\right)
+\mathrm{h.c.}
\bigg]
\bigg\}
\bigg].
\end{aligned}
\end{equation}
where the wave number is modulo $2\pi$ and ``h.c.'' denotes the Hermitian conjugate of the preceding terms inside the parentheses.
To drive Eq.~\eqref{eq:log_T_k_space}, we have used the identities
\begin{equation}
\begin{aligned}
\sum_{j=1}^n e^{i\pi j} c_j^\dagger c_{j+1}^\dagger &= \mathbf{-}\sum_{k \in K_p} e^{i k} c_k^\dagger c_{-k+\pi}^\dagger \\
&= \mathbf{-}\sum_{k \in K_p} \cos k  c_k^\dagger c_{-k+\pi}^\dagger
\end{aligned}
\end{equation}
and
\begin{equation}
\begin{aligned}
    \sum_{j=1}^n e^{i\pi j} \left( c_j^\dagger c_{j+1} + c_{j+1}^\dagger c_j \right) 
    &= \sum_{k \in K_p} \left( -e^{ik} + e^{-ik} \right) c_k^\dagger c_{k+\pi}\\
    &= -2i \sum_{k \in K_p} \sin k \, c_k^\dagger c_{k+\pi}.
\end{aligned}
\end{equation}

\subsection{BdG form for brick-wall circuits}
\label{app:sec_BdG_brick_wall}
We transform the operator in Eq.~\eqref{eq:log_T_k_space} into the BdG form. 
We define the parameter $\alpha$ by the relation 
\begin{equation}\frac{2d}{d^2+1} = \cos\alpha.
\end{equation}
We then have that \begin{equation}\frac{(d-1)^2}{d^2+1} = 1-\cos\alpha
\end{equation}
and $e_{\mathrm{H}} = \sin\alpha$. The parameters $A$ and $B$ can be written as $A=(n/2)(1-\cos \alpha)$ and $B=\cos \alpha$.
The operators $(1/x_u)\log T_a$ for $a \in \{0,1\}$ are equal to 
\begin{widetext}
\begin{equation}
\label{eq:log_T_exact_start}
\frac{n}{2}(1-\mathsf c_\alpha)
+
\sum_{k\in K_p}
\left\{
(\mathsf c_\alpha-\mathsf c_k)c_k^\dagger c_k
+
\left[
-\frac{i}{2}\mathsf s_\alpha \mathsf s_k\,
c_k^\dagger c_{-k}^\dagger
+
(-1)^a
\left(
\frac{i}{2}\mathsf s_k\,
c_k^\dagger c_{k+\pi}
+
\frac{1}{2}\mathsf s_\alpha \mathsf c_k\,
c_k^\dagger c_{-k+\pi}^\dagger
\right)
+\mathrm{h.c.}
\right]
\right\},
\end{equation}
\end{widetext}
where we have introduced the abbreviations
\begin{equation}
\begin{aligned}
    \mathsf c_k &\coloneqq \cos k, \\
    \mathsf s_k &\coloneqq \sin k, \\
    \mathsf c_\alpha &\coloneqq \cos\alpha, \\
    \mathsf s_\alpha &\coloneqq \sin\alpha .
\end{aligned}
\end{equation}
In the following, we derive the BdG form for the boundary sectors $k=0, \pi/2$ and bulk sectors for $0<k<\pi/2$.

\subsubsection{The $k=0$ sector}
If $0 \in K_p$, $k=0$ mode interacts with $k=\pi$ mode, and the relevant operator in Eq.~\eqref{eq:log_T_exact_start} is
\begin{equation}
\label{eq:O0_term}
O_0 \coloneq (\mathsf c_\alpha-1)c_0^\dagger c_0
+
(\mathsf c_\alpha+1)c_\pi^\dagger c_\pi
+
(-1)^a\mathsf s_\alpha
\left(
c_0^\dagger c_\pi^\dagger
+
c_\pi c_0
\right).
\end{equation}
We define the 2-component spinor $\Psi_{\mathrm{B},0}  \coloneq (c_0, c_\pi^\dagger)^T$ and
\begin{equation}
\begin{aligned}
    {\mathcal{H}}_a(0) & \coloneqq  x_u 
    \begin{pmatrix}
        \mathsf c_\alpha - 1 & (-1)^a \mathsf s_\alpha \\
        (-1)^a \mathsf s_\alpha & -\mathsf c_\alpha - 1
    \end{pmatrix}.
\end{aligned}
\end{equation}
The eigenvalues of ${\mathcal{H}}_a(0) $ are $0$ and $-2x_u$.
Using the canonical anticommutation relation $c_\pi^\dagger c_\pi = 1 - c_\pi c_\pi^\dagger$, we find that the BdG form
\begin{equation}
    x_uO_0 = \Psi_{\mathrm{B},0} ^\dagger {\mathcal{H}}_a(0) \Psi_{\mathrm{B},0}  + x_u(\mathsf c_\alpha + 1)
\end{equation}
is the contribution to $\log T_a$ for $k=0,\pi$.
The eigenvalues $\lambda_{0,1}$ and  $\lambda_{0,2}$ in Eq.~\eqref{eq:lambda_k=0_pi/2} are obtained by diagonalizing the $2 \times 2$ matrix  $\exp(\tilde{\mathcal{H}}_1(0))\exp(\tilde{\mathcal{H}}_0(0))$.

\subsubsection{Bulk $k$ sectors}
For $0<k<\pi/2$, the modes $k$, $-k$, $k+\pi$, and $-k+\pi$
form a closed sector. By defining
\[
\mathcal Q_k\coloneqq\{k,-k,k+\pi,-k+\pi\},
\]
where all momenta are modulo $2\pi$, the corresponding
contribution in Eq.~\eqref{eq:log_T_exact_start} is
\begin{equation}
\label{eq:Ok_compact}
\begin{aligned}
O_k
\coloneqq &
\sum_{q\in\mathcal Q_k}
\biggl\{
(\mathsf c_\alpha-\mathsf c_q)c_q^\dagger c_q
+
\biggl[
-\frac{i}{2}\mathsf s_\alpha\mathsf s_q\,
c_q^\dagger c_{-q}^\dagger
\\
&
+(-1)^a
\left(
\frac{i}{2}\mathsf s_q\,
c_q^\dagger c_{q+\pi}
+
\frac{1}{2}\mathsf s_\alpha\mathsf c_q\,
c_q^\dagger c_{-q+\pi}^\dagger
\right)
+\mathrm{h.c.}
\biggr]
\biggr\}.
\end{aligned}
\end{equation}
Moreover, for $0<k<\pi/2$, we define $(1/x_u)\mathcal H_a(k)$ by the matrix
\begin{equation}
\label{eq:BdG_Hamiltonian_brick-wall_app}
\begin{aligned}
&
\begin{pmatrix}
\mathsf c_\alpha-\mathsf c_k
&
-i\mathsf s_\alpha\mathsf s_k
&
i(-1)^a\mathsf s_k
&
(-1)^a\mathsf s_\alpha\mathsf c_k
\\
i\mathsf s_\alpha\mathsf s_k
&
-\mathsf c_\alpha+\mathsf c_k
&
-(-1)^a\mathsf s_\alpha\mathsf c_k
&
-i(-1)^a\mathsf s_k
\\
-i(-1)^a\mathsf s_k
&
-(-1)^a\mathsf s_\alpha\mathsf c_k
&
\mathsf c_\alpha+\mathsf c_k
&
i\mathsf s_\alpha\mathsf s_k
\\
(-1)^a\mathsf s_\alpha\mathsf c_k
&
i(-1)^a\mathsf s_k
&
-i\mathsf s_\alpha\mathsf s_k
&
-\mathsf c_\alpha-\mathsf c_k
\end{pmatrix}
\\
&=
\mathsf c_\alpha\, \mathbb I Z
-
\mathsf c_k\, ZZ
+
\mathsf s_\alpha\mathsf s_k\, ZY
-
(-1)^a
\big(
\mathsf s_k\, YZ
+
\mathsf s_\alpha\mathsf c_k\, YY
\big),
\end{aligned}
\end{equation}
and
\begin{equation} \label{def:nambu_spinor_app}
    \Psi_{\mathrm{B},k} = ( c_k, c_{-k}^\dagger, c_{k+\pi}, c_{-k+\pi}^\dagger )^T.
\end{equation}
The similar calculation with the $k=0$ sector yields that
\begin{equation}
    x_uO_k = \Psi_{\mathrm{B},k}^\dagger \mathcal{H}_a(k) \Psi_{\mathrm{B},k} +2 x_u \mathsf c_\alpha
\end{equation}
is the contribution to $\log T_a$ at $Q_k = \{k,-k,k+\pi,-k+\pi\}$.

\subsubsection{The $k=\pi/2$ sector}
If $\pi/2 \in K_p$, the operator of the sum of the terms involving $\pi/2, -\pi/2$ from Eq.~\eqref{eq:log_T_exact_start} is
\begin{equation}
\begin{aligned}
O_{{\pi}/{2}}
\coloneqq\;
&\mathsf c_\alpha
\left(
c_{\pi/2}^\dagger c_{\pi/2}
+
c_{-\pi/2}^\dagger c_{-\pi/2}
\right)
\\
&+
\left[
-i\mathsf s_\alpha\,
c_{\pi/2}^\dagger c_{-\pi/2}^\dagger
+
i(-1)^a
c_{\pi/2}^\dagger c_{-\pi/2}
+\mathrm{h.c.}
\right].
\end{aligned}
\end{equation}
For the spinor $\Psi_{\mathrm{B},\pi/2} = ( c_{\pi/2}, c_{-\pi/2}^\dagger, c_{-\pi/2}, c_{\pi/2}^\dagger )^T$, we obtain that  
\begin{equation} \label{eq:operator_contribution_pi/2}
    x_u O_{\pi/2} = \frac{1}{2} \Psi_{\mathrm{B},\pi/2}^\dagger \mathcal{H}_a(\pi/2) \Psi_{\mathrm{B},\pi/2} + x_u\mathsf c_\alpha,
\end{equation}
where the factor $1/2$ on the right-hand side compensates for the double counting.

Let $N_+$ be the number of the wave numbers in $\widetilde{K}_p^+$. It satisfies the equality $n = 4 N_+ + 2 \delta_{0}$, where $\delta_{0}=1$ if $0 \in K^+_p$ and $\delta_{0}=0$ otherwise.
By summing the $c_\alpha $ constants in $O_0$ and $O_k$, for $k \in \widetilde{K}^+_p,$ we have
\begin{equation}
   c_\alpha  \left( 2 N_+ + \delta_{0 \in K_p} \right) =  \frac{n}{2} c_\alpha ,
\end{equation}
which is cancelled out with the $c_\alpha $ term in $({n}/{2}) (1-c_\alpha )$ in Eq.~\eqref{eq:log_T_exact_start}, resulting in ${n}/{2}$.

Putting altogether, the operators $\log T_a$ can be written in the BdG form  as
\begin{equation}
\label{eq:BdG_form_brickwall_single_layer}
\begin{aligned}
\log T_a
&=
\frac{n}{2}x_u
+
\sum_{k\in \widetilde{K}_p^+}
w_k\,
\Psi_{\mathrm{B},k}^\dagger
\mathcal H_a(k)
\Psi_{\mathrm{B},k}
\\
&\quad
+
\delta_0
\left(
\Psi_{\mathrm{B},0} ^\dagger
{\mathcal H}_a(0)
\Psi_{\mathrm{B},0} 
+
x_u
\right),
\end{aligned}
\end{equation}
where $w_k=1$ for all $0<k<\pi/2$ and $w_{\pi/2}=1/2$.

\subsection{Block diagonalization of BdG form}
\label{app:subsec_k=pi/2}

We show that the contribution to the moment operator from $k=\pi/2$ becomes a block-diagonal form. 
The BdG Hamiltonian at $k=\pi/2$ has the following form,
\begin{equation}
    \mathcal{H}_a(\pi/2)
    =
    x_u \big[
        \mathsf c_\alpha \, \mathbb{I} \otimes Z
        + \mathsf s_\alpha \, Z \otimes Y
        - (-1)^a Y \otimes Z
    \big].
\end{equation}
Since
\begin{equation}
    \left[
        \mathsf c_\alpha \, \mathbb{I} \otimes Z
        + \mathsf s_\alpha \, Z \otimes Y,
        Y \otimes Z
    \right]
    =0,
\end{equation}
we have
\begin{equation}
\begin{aligned}
    \left[
        \mathcal{H}_0(\pi/2),
        \mathcal{H}_1(\pi/2)
    \right]
    =0 .
\end{aligned}
\end{equation}
It implies that
\begin{equation}
\begin{aligned}
    &\exp(\mathcal{H}_1(\pi/2)) \exp(\mathcal{H}_0(\pi/2))\\
    &=
    \exp\left[
        2x_u
        \left(
            \mathsf c_\alpha \, \mathbb{I} \otimes Z
            + \mathsf s_\alpha \, Z \otimes Y
        \right)
    \right],
\end{aligned}
\end{equation}
where we obtain 
\begin{equation}
\begin{aligned}
    &2x_u
    \left(
        \mathsf c_\alpha \, \mathbb{I} \otimes Z
        + \mathsf s_\alpha \, Z \otimes Y 
    \right) \\
    &=
    2x_u
    \begin{pmatrix}
        \mathsf c_\alpha & -i\mathsf s_\alpha & 0 & 0 \\
        i\mathsf s_\alpha & -\mathsf c_\alpha & 0 & 0 \\
        0 & 0 & \mathsf c_\alpha & i\mathsf s_\alpha \\
        0 & 0 & -i\mathsf s_\alpha & -\mathsf c_\alpha
    \end{pmatrix}.
    \end{aligned}
\end{equation}
Therefore, its exponential matrix is also a block-diagonal matrix:
\begin{equation}
\begin{aligned}
    &\exp(\mathcal{H}_1(\pi/2)) \exp(\mathcal{H}_0(\pi/2))\\
    &= 
    \begin{pmatrix}
        \exp(2x_u M_\alpha) & O \\
        O & \exp(2x_u M_\alpha^*)
    \end{pmatrix},
\end{aligned}
\end{equation}
where the $2 \times 2$ matrix $M_\alpha$ is defined as
\begin{equation}
    M_\alpha
    =
    \begin{pmatrix}
        \mathsf c_\alpha & -i\mathsf s_\alpha \\
        i\mathsf s_\alpha & -\mathsf c_\alpha
    \end{pmatrix}.
\end{equation}
The eigenvalues of $M_\alpha$ are $\pm1$, and therefore those of the second-moment operator at $k=\pi/2$ are $e^{2x_u}$ and $e^{-2x_u}$.

\subsection{Eigenvectors}
\label{subsection:eigenvectors}
The right and left eigenstates for given occupation numbers $\bm{n} = \{n_{k,j} \in \{0, 1\}\}_{k,j}$ are constructed by 
\begin{equation} \label{eq:biorthogonal_eigenstates}
\begin{aligned}
    \ket{R_{\bm{n}}} \coloneq& \left( \prod_{k \in \bar{K}_p^+} \prod_{j=1}^4 (\bar{\eta}_{k,j})^{n_{k,j}} \right)
    \left( \prod_{j=1}^2 (\bar{\eta}_{0,j})^{\delta_0n_{0,j}}  \right)\\
    & \cdot \left( \prod_{j=1}^2 (\bar{\eta}_{\frac{\pi}{2},j})^{\delta_\frac{\pi}{2}n_{\frac{\pi}{2},j}}  \right)
    |R_0\rangle, \\
    \bra{L_{\bm{n}}} \coloneq& \langle L_0| 
    \left( \prod_{j=1}^2 ({\eta}_{\frac{\pi}{2},j})^{\delta_\frac{\pi}{2}n_{\frac{\pi}{2},j}}  \right)\\
    & \cdot \left( \prod_{j=1}^2 ({\eta}_{0,j})^{\delta_0n_{0,j}}  \right)
    \left( \prod_{k \in \bar{K}_p^+} \prod_{j=1}^4 (\eta_{k,j})^{n_{k,j}} \right).
\end{aligned}
\end{equation}
Acting $M_{\mathrm{B},u}$ on a right eigenstate $\ket{R_{\bm{n}}} $ yields the corresponding eigenvalue
\begin{equation}
\begin{aligned}
    e^{x_u\left(n+2\delta_{0}\right)}\left( \prod_{k \in \bar{K}_p^+} \prod_{j=1}^4 \lambda_{k,j}^{n_{k,j}}\right)
     \prod_{j=1}^2 \lambda_{0,j}^{\delta_0 n_{0,j}}
     \prod_{j=1}^2 \lambda_{\frac{\pi}{2},j}^{\delta_{\frac{\pi}{2}}n_{\frac{\pi}{2},j}}.
\end{aligned}
\end{equation}
Since $\eta_{k,j}$ is a linear combination of $c_k$, $c_{-k}^\dagger$, $c_{k+\pi}$, $c_{\pi-k}^\dagger$, the vacuum for $\eta$-fermions is given by 
\begin{equation}
\ket{R_0}=\left( \prod_{k \in \bar{K}_p^+} c^{\dagger}_{-k} c^{\dagger}_{\pi-k} \right)
\left(c_{-\frac{\pi}{2}}^{\dagger} \right)^{\delta_{\frac{\pi}{2}}}
\left( c_0^{\dagger} \right)^{\delta_{0}} \ket{\textrm{vac}}_{\eta, p}.
\end{equation}
From Eq.~\eqref{eq:log_T_k_space},  the vector $\ket{R_0}$ is an eigenvector of the moment operator $M_{\mathrm{B},u}$ with eigenvalue $e^{x_u\left(n+2\delta_{0}\right)}$.

\subsection{Eigenvalues}
\label{app:eigenvalues}

By solving Eqs.~\eqref{eq:trace_M_k} and \eqref{eq:trace_I1}, we find that, for $j \in \{0,1\}$,
\begin{equation}
\label{eq:complex_bar_mu_j}
\begin{aligned}
    \bar{\mu}_{j}
    &=
    -\bigl(\cosh(\tilde{x}_{k}) - 1\bigr)\mathsf c_\alpha \mathsf c_k \\
    &\quad
    + (-1)^j
    \sqrt{
        \bigl(\cosh(\tilde{x}_{k}) + 1\bigr)
        \left[
            2
            + \bigl(\cosh(\tilde{x}_{k}) - 1\bigr)
            \mathsf c_\alpha^2 \mathsf c_k^2
        \right]
    }.
\end{aligned}
\end{equation}
The first term in Eq.~\eqref{eq:complex_bar_mu_j} becomes
\begin{equation}
\label{eq:mu_term1}
\begin{aligned}
    -\bigl(\cosh(\tilde{x}_{k})-1\bigr)\mathsf c_\alpha \mathsf c_k
    &=
    -2\sinh^2\left(\frac{\tilde{x}_k}{2}\right)
    \mathsf c_\alpha \mathsf c_k .
\end{aligned}
\end{equation}
The expression inside the square root in Eq.~\eqref{eq:complex_bar_mu_j} becomes
\begin{equation}
\label{eq:mu_term2}
\begin{aligned}
    &\bigl(\cosh(\tilde{x}_{k}) + 1\bigr)
    \left[
        2
        + \bigl(\cosh(\tilde{x}_{k}) - 1\bigr)
        \mathsf c_\alpha^2 \mathsf c_k^2
    \right] \\
    &=
    \sinh^2(\tilde{x}_{k})\mathsf c_\alpha^2 \mathsf c_k^2
    + 2\bigl(\cosh(\tilde{x}_{k})+1\bigr) \\
    &=
    4\cosh^2\left(\frac{\tilde{x}_k}{2}\right)
    \left[
        \left(
            \sinh\left(\frac{\tilde{x}_k}{2}\right)
            \mathsf c_\alpha \mathsf c_k
        \right)^2
        + 1
    \right].
\end{aligned}
\end{equation}
By the notation $ q_k=
    \sinh\left({\tilde{x}_k}/{2}\right)\mathsf c_\alpha \mathsf c_k$, 
we rewrite Eq.~\eqref{eq:complex_bar_mu_j} as
\begin{equation}
\label{eq:bar_mu_half_angle}
\begin{aligned}
    \frac{\bar{\mu}_{j}}{2}
    =
    -q_k\sinh\left(\frac{\tilde{x}_k}{2}\right)
    + (-1)^j
    \cosh\left(\frac{\tilde{x}_k}{2}\right)
    \sqrt{q_k^2+1}.
\end{aligned}
\end{equation}
Equivalently, by the notation $\theta_k = \operatorname{arsinh} q_k ,$
we have
\begin{equation}
\begin{aligned}
    \frac{\bar{\mu}_0}{2}
    &=
    \cosh\left(\frac{\tilde{x}_k}{2}-\theta_k\right), \\
    \frac{\bar{\mu}_1}{2}
    &=
    -\cosh\left(\frac{\tilde{x}_k}{2}+\theta_k\right).
\end{aligned}
\end{equation}
Therefore, from $\mu_j+\mu_j^{-1}=\bar{\mu}_j$, we obtain
\begin{equation}
\label{eq:mu_j_theta}
\begin{aligned}
    \mu_0
    &=
    \exp\left[
        \frac{\tilde{x}_k}{2}-\theta_k
    \right], \\
    \mu_1
    &=
    -
    \exp\left[
        \frac{\tilde{x}_k}{2}+\theta_k
    \right].
\end{aligned}
\end{equation}
In terms of $q_k$, this is
\begin{equation}
\label{eq:mu_j_q}
\begin{aligned}
    \mu_0
    &=
    e^{\tilde{x}_k/2}
    \left(
        \sqrt{q_k^2+1}-q_k
    \right), \\
    \mu_1
    &=
    -
    e^{\tilde{x}_k/2}
    \left(
        \sqrt{q_k^2+1}+q_k
    \right).
\end{aligned}
\end{equation}

The eigenvalues $\lambda_{k,i}$ of $\exp(C_k) = \widetilde{M}_k^2$, for $k \in \bar{K}_p^+$, are obtained by the squares of $\mu_{i}$, which are
\begin{equation}
    \begin{aligned}
        \label{eq-def-lambda-k}
         \lambda_{k,1} &= e^{2({x}_u-\theta_k)}, \\
             \lambda_{k,2} &= e^{-2({x}_u-\theta_k)},\\
    \lambda_{k,3} &=e^{2({x}_u+\theta_k)}\\
    \lambda_{k,4} &= e^{-2({x}_u+\theta_k)},
    \end{aligned}
\end{equation}
where we have defined
\begin{equation}
    \theta_k \coloneq \arsinh \big(\sinh\left(x_u\right) \cos\alpha \cos k \big),
\end{equation}
where $\arsinh$ is the inverse function of $\sinh$.
We note that $\theta_k=\theta_{-k}$. Also, we have that $\lambda_{k,1}=1/\lambda_{k,2}$ and $\lambda_{k,3}=1/\lambda_{k,4}$.
For $k=0,\pi/2$, we obtain
\begin{equation} \label{eq:lambda_k=0_pi/2}
    \begin{aligned}
             \lambda_{0,1} &= e^{-2({x}_u-\theta_0)},\\
         \lambda_{0,2} &= e^{-2({x}_u+\theta_0)},\\
         \lambda_{\frac{\pi}{2},1} &= e^{2{x}_u},\\
         \lambda_{\frac{\pi}{2},2} &= e^{-2{x}_u},
    \end{aligned}
\end{equation}
where $\theta_0$ denotes $\theta_k$  evaluated at $k=0$, and for $k=\pi/2$, we relabeled $\lambda_{k,j}$ by combining two eigenvalues with the same wave number.
The eigenvalues in Eq.~\eqref{eq:lambda_k=0_pi/2} are derived by corresponding $2 \times 2$ matrices, discussed in  Appendix~\ref{app:sec_BdG_brick_wall}.
From these values, the eigenvalues of the second-moment operator $M_{\mathrm{B},u}$ are given by
\begin{equation} \label{eq:eigenvalues:moment_op_M_B_u}
    e^{x_u\left(n+2\delta_{0}\right)}\left( \prod_{k \in \bar{K}_p^+} \prod_{j=1}^4 \lambda_{k,j}^{n_{k,j}}\right)
     \prod_{j=1}^2 \lambda_{0,j}^{\delta_0 n_{0,j}}
     \prod_{j=1}^2 \lambda_{\frac{\pi}{2},j}^{\delta_{\frac{\pi}{2}}n_{\frac{\pi}{2},j}},
\end{equation}
where $\{n_{k,j}\}_{k\in K_p^+,j}$ is the set of the occupation numbers with $n_{k,j} \in \{0,1\}$, and $\sum_{k,j} n_{k,j}$ is even so that it satisfies the fermionic parity condition.

We now identify the largest eigenvalue. Since the real part of
\begin{equation}
    x_u=\frac{1}{2}\log\left(1-\frac{e_u}{e_{\mathrm H}}\right)
\end{equation}
is negative in the parameter regime considered here, the largest
eigenvalue is obtained by filling the quasi-particles with eigenvalues
$\lambda_{k,2}$ and $\lambda_{k,4}$ for all $k\in\bar K_p^+$, together
with $\lambda_{0,1}$, $\lambda_{0,2}$, and
$\lambda_{\frac{\pi}{2},2}$ when the corresponding exceptional momenta
exist. The prefactor $e^{x_u(n+2\delta_0)}$ in
Eq.~\eqref{eq:eigenvalues:moment_op_M_B_u} is exactly cancelled by the
factors $e^{-2x_u}$ coming from these filled modes. Thus the largest
eigenvalue is $1$, as expected for a moment operator.

It is useful to rewrite the spectrum relative to this filled configuration. For each $k\in\bar K_p^+$, the four-component Nambu spinor
contains the four wave numbers
\begin{equation}
    k,\quad -k,\quad \pi-k,\quad k-\pi .
\end{equation}
Filling an originally empty mode with factor $\lambda_{k,1}$ gives the
multiplier $\lambda_{k,1}$. On the other hand, creating a hole in the
filled reciprocal mode $\lambda_{k,2}$ gives the multiplier
$\lambda_{k,2}^{-1}=\lambda_{k,1}$. These two equivalent multipliers are
naturally assigned to the two wave numbers $k$ and $-k$. Similarly,
$\lambda_{k,3}$ and $\lambda_{k,4}^{-1}=\lambda_{k,3}$ are assigned to
$\pi-k$ and $k-\pi$.

Therefore, we define a quasi-energy on the set $K_p$ by
\begin{equation}
    \lambda_k \coloneq e^{2(x_u-\theta_k)},
    \qquad k\in K_p .
\end{equation}
Equivalently, for $k\in\bar K_p^+$,
\begin{equation}
    \begin{aligned}
         \lambda_k&=\lambda_{-k}
    =
    \lambda_{k,1}
    =
    \lambda_{k,2}^{-1},\\
    \lambda_{\pi-k}&=\lambda_{k-\pi}
    =
    \lambda_{k,3}
    =
    \lambda_{k,4}^{-1}.
    \end{aligned}
\end{equation}
For the exceptional wave numbers, the same notation gives
\begin{equation}
\begin{aligned}
    \lambda_0&=\lambda_{0,1}^{-1},\\
    \lambda_\pi&=\lambda_{0,2}^{-1},\\
    \lambda_{\frac{\pi}{2}}
    &=
    \lambda_{-\frac{\pi}{2}}
    =
    \lambda_{\frac{\pi}{2},1}
    =
    \lambda_{\frac{\pi}{2},2}^{-1}.
\end{aligned}
\end{equation}
With this notation, the eigenvalues of $M_{\mathrm{B},u}$ take the
compact form
\begin{equation}
\begin{aligned}
    \prod_{k\in K_p}\lambda_k^{n_k},
\end{aligned}
\end{equation}
where $n_k\in\{0,1\}$ denotes the occupation number of the corresponding
quasi-particle excitation and
\begin{equation}
\begin{aligned}
    \sum_{k\in K_p} n_k \in 2\mathbb{Z}.
\end{aligned}
\end{equation}
The choice $n_k=0$ for all $k\in K_p$ corresponds to the largest
eigenvalue $1$.

Finally, using $x_u
    =
    ({1}/{2})
    \log\left(1-{e_u}/{e_{\mathrm H}}\right)$ and
    $\cos\alpha
    =
    {2d}/({d^2+1}),$
we can rewrite $\lambda_k$ as
\begin{equation}
\begin{aligned}
    \lambda_k
    =
    \left(
        a_k\widetilde e_u
        +
        \sqrt{
            a_k^2\widetilde e_u^2+1-\widetilde e_u
        }
    \right)^2,
\end{aligned}
\end{equation}
where
\begin{equation}
\begin{aligned}
    \widetilde e_u
    &=
    \frac{e_u}{e_{\mathrm H}},\\
    a_k
    &=
    \frac{d}{d^2+1}\cos k .
\end{aligned}
\end{equation}

\section{Technical lemmas}
\label{app:technical_lemma}

We prove the lemmas that we have used for the technical derivation in Subsections~\ref{subsec:ecaxt_diagonalization} and \ref{sec:formula_evalues_brick_wall}.
The following is the restatement of Lemma~\ref{lemma:product_exponentials_nambu_spinors}.

\begin{lem}[Lemma 6 restated] \label{lemma:product_exponentials_nambu_spinors_app}
    Let $\Psi_{\mathrm{B},k}$ be the Nambu spinor defined in Eq.~\eqref{def:nambu_spinor_app}.
    If $4 \times 4$ matrices $A$ and $B$  satisfy the relation $e^Ae^B=e^C$ for some $C$, then the corresponding relation 
\begin{equation}
    e^{\Psi_{\mathrm{B},k}^\dagger A \Psi_{\mathrm{B},k}} e^{\Psi_{\mathrm{B},k}^\dagger B \Psi_{\mathrm{B},k}} = e^{\Psi_{\mathrm{B},k}^\dagger C \Psi_{\mathrm{B},k}}
\end{equation}
for the fermionic operators holds.
\end{lem}
\begin{proof}
    For any matrices $A$ and $B$, we define $\hat{A} \coloneq \Psi_{\mathrm{B},k}^\dagger A \Psi_{\mathrm{B},k}$ and $\hat{B} \coloneq \Psi_{\mathrm{B},k}^\dagger B \Psi_{\mathrm{B},k}$, and then we have the commutation relation 
    \begin{equation} \label{eq:commutator_nambu_spinor_app}
        [\hat{A}, \hat{B}] = \Psi_{\mathrm{B},k}^\dagger [A, B] \Psi_{\mathrm{B},k}. 
    \end{equation}
    This follows from the commutator identity $[XY, ZW] = X\{Y,Z\}W - XZ\{Y,W\} + \{X,Z\}WY - Z\{X,W\}Y$ for operators $X,Y,Z,W$, and the fermionic canonical anticommutation relations: $\{\Psi_{\mathrm{B},k,i}^\dagger, \Psi_{\mathrm{B},k,j}\} = \delta_{i,j}$ and $\{\Psi_{\mathrm{B},k,i}^\dagger, \Psi_{\mathrm{B},k,j}^\dagger\} = \{\Psi_{\mathrm{B},k,i}, \Psi_{\mathrm{B},k,j}\} = 0$. 
    By the Baker-Campbell-Hausdorff formula, we have $e^{\hat{A}} e^{\hat{B}}=e^{\hat{C}}$, where $\hat{C}$ is given by an infinite series of nested commutators:
\begin{equation} \label{eq:BCH_formula_app}
    \hat{C} = \hat{A} + \hat{B} + \frac{1}{2}[\hat{A}, \hat{B}] + \frac{1}{12}[\hat{A}, [\hat{A}, \hat{B}]] - \frac{1}{12}[\hat{B}, [\hat{A}, \hat{B}]] + \cdots
\end{equation}
Each term in Eq.~\eqref{eq:BCH_formula_app} is in the form $[\hat{X}_1, [\hat{X}_2, \dots, [\hat{X}_{m-1}, \hat{X}_m]\dots]]$, for some $\hat{X}_j \in \{\hat{A}, \hat{B}\}$. By using Eq.~\eqref{eq:commutator_nambu_spinor_app} recursively, it becomes $\Psi_{\mathrm{B},k}^\dagger[{X}_1, [{X}_2, \dots, [{X}_{m-1}, {X}_m]\dots]]\Psi_{\mathrm{B},k}$. Therefore, we obtain $\hat{C}=\Psi_{\mathrm{B},k}^\dagger C\Psi_{\mathrm{B},k}$, where $C$ satisfies $e^Ae^B=e^C$.
\end{proof}

The following is the restatement of Lemma~\ref{lem:orthogonal_reciprocal_pairs}.

\begin{lem}[Lemma 7 restated] \label{lem:orthogonal_reciprocal_pairs_app}
The $4 \times 4$ matrix $\widetilde{M}_k = Z_1 e^{{w_k} \mathcal{H}_0(k)}$, where $U=Z \otimes \mathbb{I}$ and $\mathcal{H}_a(k)$ is defined by Eq.~\eqref{eq:BdG_Hamiltonian_brick-wall_app},  satisfies $J_O \widetilde{M}_k^T J_O = \widetilde{M}_k^{-1}$, where $J_O \coloneq Z \otimes Y$. Consequently, its eigenvalues form reciprocal pairs: if $\lambda \neq 0$ is an eigenvalue, $\lambda^{-1}$ is also an eigenvalue.
\end{lem}

\begin{proof}
By taking the matrix transpose of $\mathcal{H}_0(k)$, we obtain that
\begin{equation}
\label{eq:H_transpose_app}
\begin{aligned}
    \mathcal{H}_0(k)^T
    &=
    x_u \big(
        \cos\alpha \, \mathbb{I}_2 Z
        - \cos k \, Z Z
        - \sin\alpha \sin k \, Z Y \\
    &\qquad
        + \sin k \, Y Z
        - \sin\alpha \cos k \, Y Y
    \big).
\end{aligned}
\end{equation}
It is straightforward to show that 
\begin{equation} \label{eq:H_symmetry_identity_app}
    J_O \mathcal{H}_a(k)^T J_O = -\mathcal{H}_a(k),
\end{equation}
which implies
\begin{equation}
\begin{aligned} \label{eq:orthogonal_M_app}
    J_O \widetilde{M}_k^T J_O &= \left( J_O e^{{w_k} \mathcal{H}_0(k)^T} J_O \right) U \\
    &= e^{{w_k}  J_O \mathcal{H}_0(k)^T J_O} U \\
    &= e^{-{w_k} \mathcal{H}_0(k)} U\\
    &= \widetilde{M}_k^{-1}.
\end{aligned}
\end{equation}
Here, we have used the relation $[U,J_O]=0$.

We define the characteristic polynomial as $P(\lambda) \coloneq \det(\widetilde{M}_k - \lambda \mathbb{I}_4)$. From Eq.~\eqref{eq:orthogonal_M_app} and $\det(J_O) = 1$, we have
\begin{equation}
\begin{aligned}
    P(\lambda) &= \det(\widetilde{M}_k^T - \lambda \mathbb{I}_4) \\
    &= \det(J_O (\widetilde{M}_k^{-1} - \lambda \mathbb{I}_4) J_O) \\
    &= \det(\widetilde{M}_k^{-1} - \lambda \mathbb{I}_4) \\
    &= \det\left( \widetilde{M}_k^{-1} (\mathbb{I}_4 - \lambda \widetilde{M}_k) \right) \\
    &= \det(\widetilde{M}_k)^{-1} \lambda^4 \det(\widetilde{M}_k - \lambda^{-1} \mathbb{I}_4).
\end{aligned}
\end{equation}
Because $\det(U)=1$ and $\Tr(H_0(k))=0$, it holds that \begin{equation}\det(\widetilde{M}_k) = 1,
\end{equation}and we obtain $P(\lambda) = \lambda^4 P(\lambda^{-1})$. This guarantees that if $\lambda \neq 0$ is an eigenvalue, $\lambda^{-1}$ is also an eigenvalue.
\end{proof}

\section{Frame potential}
\label{sec-frame-potential}

The $k$-th frame potential \cite{gross2007evenly} of random circuits $\nu$ with the circuit depth $t$ is defined by
\begin{equation} \label{eq:frame_potential}
    F^{(k)}_{\nu_t} = \int |\textrm{Tr}(U^{\dagger}V)|^{2k} d\nu_t(U)d\nu_t(V).
\end{equation}
Equivalently, we can rewrite it in terms of the moment operator as
\begin{equation} 
\label{eq:frame_potential2}
F^{(k)}_{\nu_t} =\textrm{Tr}\left[ \left( M_{\nu}^{(k)} \right)^t \cdot \left( {M_{\nu}^{(k)}}^{\dagger} \right)^t \right].
\end{equation}
From our diagonalization results in the main text, we can obtain the exact value of the second frame potential in principle, since we have obtained all eigenvalues and eigenvectors.
However, the exact formula for the frame potential involves $t$-th power summation of eigenvalues, and it might not be concise. 
In this section, we obtain the upper-bound on the frame potential of $u$-structured brick-wall circuits with slight modification, by the standard technique of mapping to partition functions of Ising-like model \cite{PhysRevX.8.021014, hunter2019unitary}.

\subsection{Second-order frame potential}

\begin{figure*}
    \centering 
    \includegraphics[width=130mm]{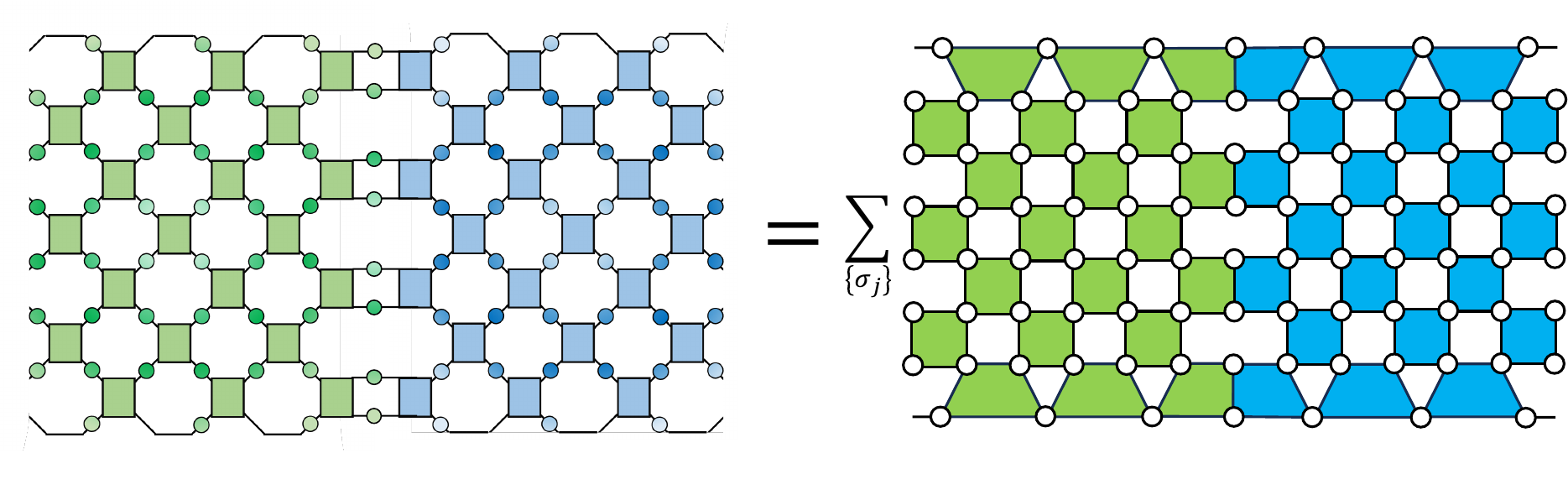}
    \caption{The frame potential (left-hand side) and the classical spin model (right-hand side).}\label{fig:QC_statmech}
\end{figure*}

We consider the case of $k=2$. The frame potential of the ensemble of local random circuits can be mapped to the partition function of a corresponding Ising model. This mapping has been studied for 
random circuits with Haar random two-qudit gates to calculate the entanglement entropy \cite{zhou2019emergent} and the frame potential \cite{hunter2019unitary}.
We average each single-qudit gate over Haar measure in Eq.~\eqref{eq:frame_potential}, it is equal to the partition function of the classical statistical mechanics model consisting of $k$-state spins (Fig. \ref{fig:QC_statmech}), where the states are $\{ \sigma_i \}_{i=1, \dots, k}$, and the four-body Boltzmann weights are obtained by
\begin{align}
     W^u(\sigma_1, \sigma_2 ,\sigma_3 ,\sigma_4)=\big(\bra{\sigma_3} \otimes \bra{\sigma_4} \big)u^{\otimes 2,2} \big( \ket{\tilde{\sigma_1}} \otimes \ket{\tilde{\sigma_2}} \big),
\end{align}
where $\sigma\in \{I,S\}$ and $\ket{\tilde{\sigma_i}}$ for $i=1,2$ are given by Eq.~\eqref{eq-def-Itilde-Stilde-states}.
Graphically, the weights are 
\begin{equation}
\begin{aligned} \label{bw_general}
    \vcenter{\hbox{\includegraphics[width=19mm]{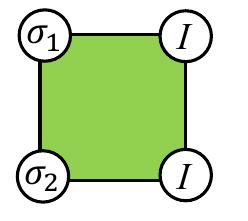}}}=& \,
    \begin{cases}
        1 &  (\sigma_1=\sigma_2=I),\\
        0 & \textrm{otherwise},
    \end{cases}\\
\vcenter{\hbox{\includegraphics[width=17mm]{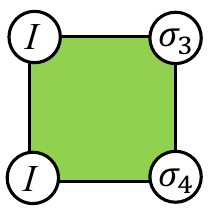}}}
=& \, 
\frac{1}{(d^2-1)^2}
\Big(
d^3+d
-\frac{\bra{\sigma_3,\sigma_4} u^{\otimes 2,2} \ket{S,I}}{d}\\
&-\frac{\bra{\sigma_3,\sigma_4} u^{\otimes 2,2} \ket{I,S}}{d}
\Big),
\\
\vcenter{\hbox{\includegraphics[width=17mm]{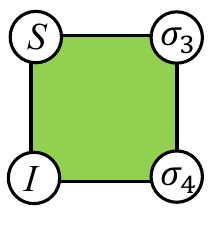}}}
=& \, 
\frac{1}{(d^2-1)^2}
\Big(
-2d^2
+\bra{\sigma_3,\sigma_4} u^{\otimes 2,2} \ket{S,I}\\
&+\frac{\bra{\sigma_3,\sigma_4} u^{\otimes 2,2} \ket{I,S}}{d^2}
\Big).
\end{aligned}
\end{equation}
for $\sigma_3 \neq \sigma_4$. Other weights are obtained by the spin-flip symmetry $W(\sigma_1 ,\sigma_2 ,\sigma_3 ,\sigma_4)=W(\bar{\sigma_1}, \bar{\sigma_2} ,\bar{\sigma_3} ,\bar{\sigma_4})$, where $\bar{I}=S$ and $\bar{S}=I$.

\subsection{Frame potential in single-domain wall sector}
We consider a structured circuit in which we apply two distinct layers alternatively. 
Specifically, we consider an analytically tractable case where the interaction in the layers at an odd time (blue triangles in Eq.~\eqref{eq-def-trans-mat-fp}) satisfies the 2-design condition, that is, $W^u(I,S,I,S)=W^u(I,S,S,I)=0$, and the interactions in the layers at even time can be arbitrary. 
We use the notation $a=W^u(I,I,S,I)$, $b=W^u(I,S,I,S)$, $c=W^u(I,S,S,I)$ for the Boltzmann weight in the layers at even time (green boxes in Eq.~\eqref{eq-def-trans-mat-fp}). In terms of the entangling power $e_u$ and the gate typicality $g_u$, we have 
\begin{equation}
    \begin{aligned}
        a=&\frac{d}{d^2-1}e_u,\\
        b=&1-\frac{d^2+1}{2(d^2-1)}e_u-g_u,\\
        c=&-\frac{d^2+1}{2(d^2-1)}e_u + g_u.
    \end{aligned}
\end{equation}
To exclude trivial examples, we assume that $u$ is not locally equivalent to the identity gate, namely, $e_u\ne 0$ and $g_u \ne 0$.
The solvability condition is equivalent to $c=0$.

For technical simplicity, we apply the layer of Haar unitaries at the end of the circuit.
Then, the frame potential can be written in terms of the transfer matrix $T$ as follows:
\begin{align} \label{eq-def-trans-mat-fp}
\includegraphics[width=80mm]{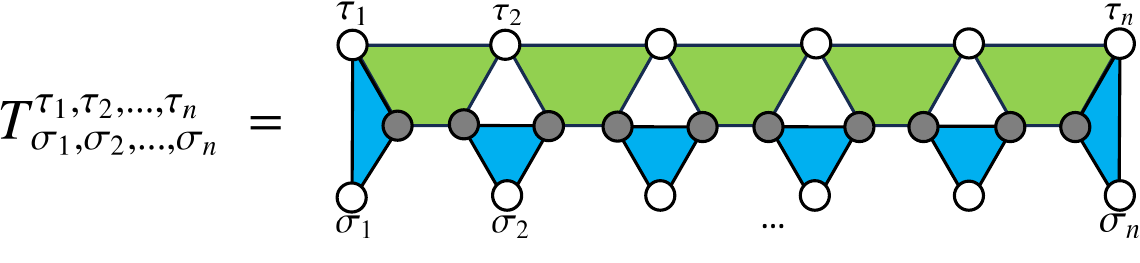},
\end{align}
where we have rotated the expression by 90 degrees, and the spins (balls) with gray color are summed over, $\sigma_i, \tau_i \in \{I,S\}$, and we obtain
\begin{align}
    F^{(2)}=\textrm{Tr}(T^{2t}).
\end{align}
Now, the problem of computing the frame potential reduces to computing the eigenvalues of the transfer matrix $T$.
The leading eigenvalues 
are 1 and the corresponding eigenvectors are zero domain-wall configurations, namely,
\begin{align}
    &T\ket{I}^{\otimes n}=\ket{I}^{\otimes n},\\
    &T\ket{S}^{\otimes n}=\ket{S}^{\otimes n}.
\end{align}

To obtain the next-leading terms, we consider the transfer matrix $T$ restricted to the subspace spanned by single domain-wall sector $S_1=\textrm{span}\{ \ket{i} \}_{i=1}^{n-1}$, where we have defined $\ket{i}=\ket{I}^{\otimes i}\otimes \ket{S}^{\otimes n-i}$, and we diagonalize the matrix in the subspace. We note that the diagonalization for Haar random brick-wall circuits is obtained in Ref.~\cite{rampp2024hayden}, and we extend the result to structured random brick-wall circuits.  In the subspace, the $I$-type domain is always left side of the $S$-type domain, and the other subspace, where the $S$-type domain is left to the $I$-type can be treated in the same way. The diagonal element $\bra{i}T\ket{i}$ can be obtained by summing the four domain wall configurations as follows:
\begin{widetext}
    \begin{align}
    \includegraphics[width=110mm]{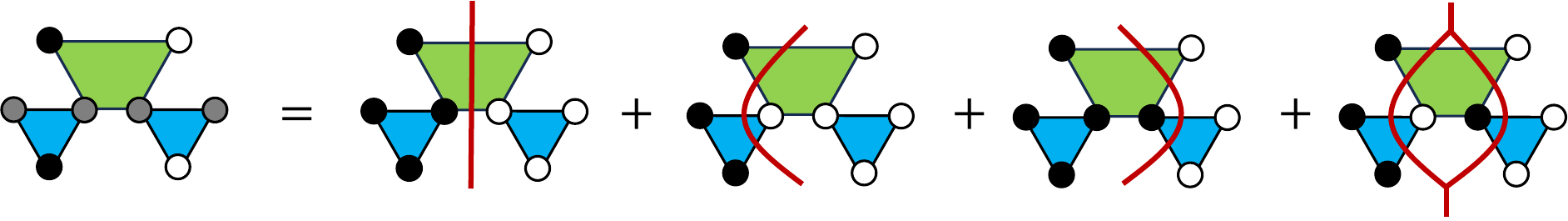},
\end{align}
\end{widetext}
where the black ball is $I$ and the white ball $S$, and it implies 
\begin{equation}
\begin{aligned}
\bra{i}T\ket{i}
&=
b
+2\frac{d}{d^2+1}a
+\left(\frac{d}{d^2+1}\right)^2 c
\\
&=
1
-\frac{d^4-d^2+1}{2(d^4-1)}\,e_u
-\frac{d^4+d^2+1}{(d^2+1)^2}\,g_u .
\end{aligned}
\end{equation}
where in the first equality, we have used the equality $W_{\mathrm{H}}(I,I,S,I)={d}/{(d^2+1)}$.
The off-diagonal elements in the single-domain wall subspace are nonzero only for $\bra{i}T\ket{i+1}$ and $\bra{i+1}T\ket{i}$ because a domain wall can move at most one site by one transfer matrix. Also, we have $\bra{i}T\ket{i+1}=\bra{i+1}T\ket{i}$. They are a sum of the two configurations as
\begin{align}
    \includegraphics[width=80mm]{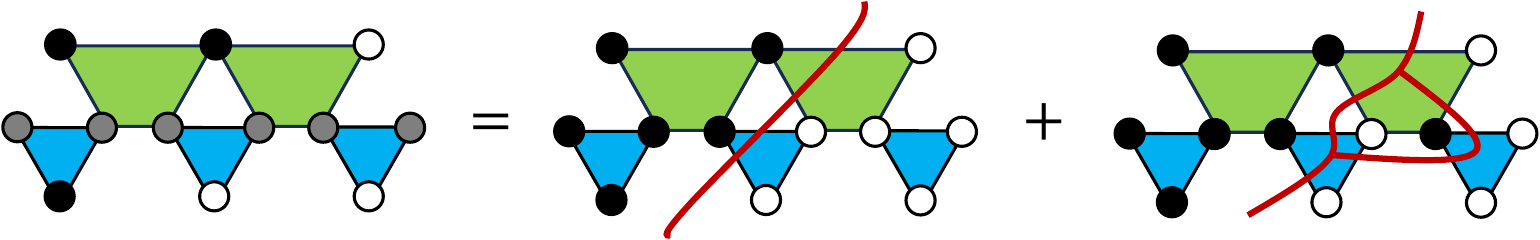},
\end{align}
which implies that $\bra{i}T\ket{i+1}$ is equal to 
\begin{equation}
\begin{aligned}
&\frac{d}{d^2+1}a+
\left(\frac{d}{d^2+1}\right)^2c\\
&=\frac{d^2}{2(d^2+1)^2(d^2-1)}
\left[
(d^2+1)e_u+2(d^2-1)g_u
\right].
\end{aligned}
\end{equation}

For most parameter regimes, the single domain-wall subspace is not an invariant subspace of $T$. Indeed, unless $a=0$ (where the two-qudit interaction $V$ is either identity or $\textrm{SWAP}$ up to single-qudit unitaries), the number of domain walls can increase by one at a boundary and increase by two in the bulk as the example configurations
\begin{align}
    \includegraphics[width=22mm]{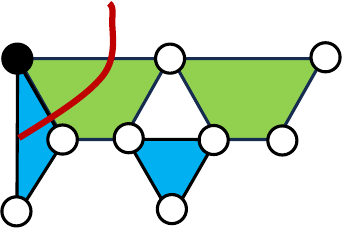}, \ \ \ \ \ 
    \includegraphics[width=35mm]{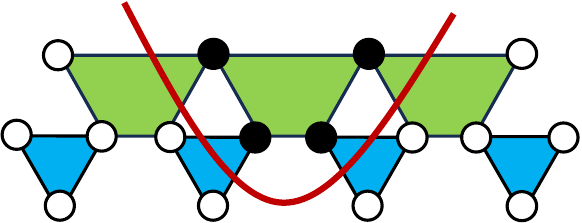}.
\end{align}
Moreover, if there is more than one domain wall, unless $c=0$, the number of domain walls can decrease at a boundary, for example,
\begin{align}
    \includegraphics[width=22mm]{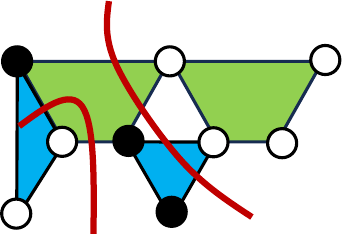}.
\end{align}

We first consider the solvable condition $c=0$ and later discuss the general case. 
When $c=0$, the number of domain walls can only increase, but such configurations do not contribute to the frame potential. This is because the frame potential is the trace of $T^{2t}$, and the initial and final spin configurations have to be the same, and therefore the number of domain wall is the same for all time. 
Then, it suffices to consider $T$ restricted to the single domain wall subspace to obtain the leading eigenvalues.
Let $P_1$ be the projector onto the subspace $S_1$.
Then, we have that 
\begin{equation}
\begin{aligned}
   P_1TP_1= &\left(1-\frac{d^4+1}{d^4-1}e_u \right) \sum_{i=1}^{n-1}\ket{i}\bra{i}\\
    &+\frac{d^2}{d^4-1}e_u \sum_{i=1}^{n-2}\left(\ket{i}\bra{i+1}+\ket{i+1}\bra{i}\right).
\end{aligned}
\end{equation}
The last equation above is a Hamiltonian of the tight-binding model with the open boundary condition, and for $k=1,2, \dots, n-1$, the eigenvalues $\lambda_k$ and eigenvectors $\ket{\phi_k}$ are 
\begin{align}
\lambda_k&=
1-\frac{d^4+1}{d^4-1}e_u +\frac{d^2}{d^4-1}e_u \cos{\frac{k \pi}{n}}, \\
\ket{\phi_k}&=\sum_{i=1}^{n-1}\sin{\frac{ki\pi}{n}\ket{i}}.
\end{align}
The second-largest eigenvalue is $1-\frac{d^4+1}{d^4-1}e_u+\frac{d^2}{d^4-1}e_u \cos{\frac{ \pi}{n}}$, and it becomes small when $e_u$ is large. This is a rough reason why large $e_u$ makes the frame potential smaller than the case of Haar random gates.
For the case of Haar random gates, the eigenvalues are not the same as those in Theorem~\ref{thm:exact_spectrum_local}, because in this section we consider the open boundary condition.

\begin{figure*}[t]
    \centering
    \includegraphics[width=100mm]{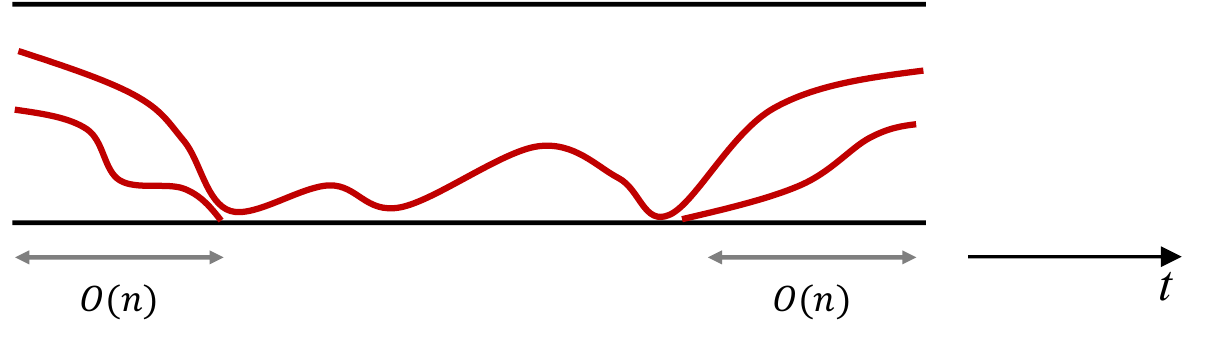}
    \caption{Non-conservation of the number of domain walls.}
    \label{fig:nonzero_c_DW_decrease}
\end{figure*}

We define 
\begin{equation}
x:=1-\frac{d^4+1}{d^4-1}e_u
\end{equation}
and \begin{equation}
y:=\frac{d^2}{d^4-1}e_u.
\end{equation}
The contribution of single domain-wall configurations to the frame potential, which we denote by $f_1$, is 
\begin{equation}
\begin{aligned}
f_1
&=
\sum_{k=1}^{n-1}\lambda_k^{2t}
=
\sum_{k=1}^{n-1}
\left(
x+y\cos\frac{k\pi}{n}
\right)^{2t}
\\
&=
\sum_{\ell=0}^{t}
\binom{2t}{2\ell}
x^{2t-2\ell}y^{2\ell}
\sum_{k=1}^{n-1}
\cos^{2\ell}\frac{k\pi}{n}
\\
&=
\sum_{\ell=0}^{t}
\binom{2t}{2\ell}
x^{2t-2\ell}y^{2\ell}
\left[
\frac{n}{4^\ell}
\sum_{m=-\lfloor \ell/n \rfloor}^{\lfloor \ell/n \rfloor}
\binom{2\ell}{\ell+mn}
-1
\right]
\\
&=
n
\sum_{m=-\lfloor t/n \rfloor}^{\lfloor t/n \rfloor}
\sum_{\ell=|m|n}^{t}
\binom{2t}{2\ell}
\binom{2\ell}{\ell+mn}
\frac{x^{2t-2\ell}y^{2\ell}}{4^\ell}
\\
&\quad
-\frac{1}{2}
\left[
(x+y)^{2t}+(x-y)^{2t}
\right].
\end{aligned}
\label{eq:f1_random_walk}
\end{equation}
where we have used the equalities $\sum_{k=1}^{n-1}\cos^{2i+1}\left( \frac{k\pi}{n} \right)=0$ and $\sum_{k=0}^{n-1}\cos^{2i}\left( \frac{k\pi}{n} \right)=\frac{n}{4^i}\sum_{k=-\lfloor i/n \rfloor}^{k= \lfloor i/n \rfloor}\binom{2i}{i+kn}$ for an integer $i$. Moreover, we have that 
\begin{widetext}
\begin{equation}
\begin{aligned}
f_1
&=
n x^{2t}
\sum_{m=-\lfloor t/n \rfloor}^{\lfloor t/n \rfloor}
\binom{2t}{2|m|n}
\left(\frac{y}{2x}\right)^{2|m|n}
\pFq{2}{1}
{|m|n-t,\ |m|n-t+\frac{1}{2}}
{1+2|m|n}
{\frac{y^2}{x^2}}
-\frac{(x+y)^{2t}+(x-y)^{2t}}{2}.
\end{aligned}
\end{equation}
\end{widetext}
where we have used the hypergeometric function,
\begin{equation}
\begin{aligned}
\pFq{2}{1}{a,b}{c}{z}
&=
\sum_{i=0}^{\infty}
\frac{(a)_i(b)_i}{(c)_i}
\frac{z^i}{i!},
\\
(a)_i
&\coloneqq
\prod_{j=0}^{i-1}(a+j).
\end{aligned}
\end{equation}
These single domain-wall configurations contribute dominantly to the frame potential, and from the above expression, we find that the contributions become small when $e_u$ is large.

Finally, we give a non-rigorous but intuitive argument for the value of the frame potential for non-zero $c$, and we argue that the 
single-domain calculation is not a good approximation for $c \neq 0$ for $\Omega(n)$-depth.
When $c \ne 0$, spin configurations without conserving the number of domain walls can have {a} nonzero value. As we discuss 
below, if $t$ is much greater than $n$, violation of the conservation of the number of the domain wall should increase the frame potential compared to the case $c = 0$.
The weight of the spin configurations with domain walls roughly decays exponentially in the total length of domain walls. If there are domain walls $k$ and the number is conserved, then the total length is $\Omega(kt)$, and the weight is $O(e^{-kt})$. However, if the number does not conserve (for example, Fig.~\ref{fig:nonzero_c_DW_decrease}), the domain-wall number can be one after $O(n)$ time steps, and the dominant weight of configurations starting from $k$ domain-walls is only $O(e^{-2nk-(t-2n)})=O(e^{-t})$.

\bibliography{Ref}

\end{document}